\documentclass[aps, prx, reprint, showpacs, superscriptaddress]{revtex4-2}
\usepackage{graphicx}
\usepackage{amsmath,amsfonts,amssymb,bm,amsthm,bbm}
\usepackage{hyperref}%
\usepackage{tikz}
\usepackage{changes}

\newcommand{\rd}{\textrm{d}}
\newcommand{\re}{\textrm{e}}
\newcommand{\inp}[2]{\langle#1|#2\rangle}
\newcommand{\ket}[1]{|#1\rangle}
\newcommand{\bra}[1]{\langle#1|}
\newcommand{\ave}[1]{\langle #1 \rangle}
\makeatletter
\newcommand*{\rom}[1]{\expandafter\@slowromancap\romannumeral #1@}
\makeatother

\begin{document}
\title{Quantum Oscillation in Excitonic Insulating Electron-Hole Bilayer}

\author{Yuelin Shao}
\email{ylshao@iphy.ac.cn}
\affiliation{Beijing National Laboratory for Condensed Matter Physics and Institute of Physics, Chinese Academy of Sciences,
Beijing 100190, China}
\affiliation{School of Physical Sciences, University of Chinese Academy of Sciences, Beijing 100049, China}
\author{Xi Dai}
\email{daix@ust.hk}
\affiliation{Department of Physics, The Hongkong University of Science and Technology,
Clear Water Bay, Kowloon 999077, Hong Kong, China}

\date{\today}

\begin{abstract}

  We study the quantum oscillations of inter-layer capacitance in an excitonic insulating electron-hole double layer with the Hartree Fock mean-field theory.
  Such oscillations could be simply understood from the physical picture ``exciton formed by electron/hole Landau levels'', where the direct gap between the electron-hole Landau levels will oscillate with exciton chemical potential and the inverse of the magnetic field.
  We also find that the excitonic order parameters can be destroyed by a strong magnetic field.
  At this time, the system becomes two independent quantum Hall liquids and the inter-layer capacitance oscillates to zero at zero temperature.

\end{abstract}
\maketitle

\section{Introduction}

\begin{figure}
  \centering
  \includegraphics[width=\linewidth]{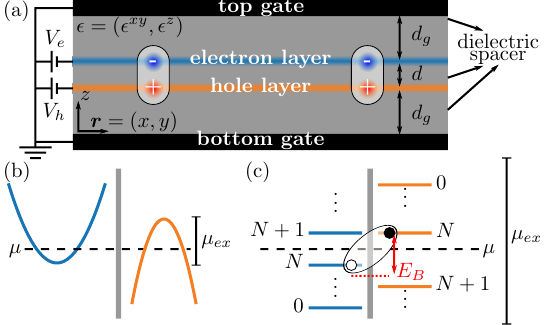}
  \caption{(a) Setup of the double-gated electron-hole bilayer system. $d$ is the geometry distance between the electron-hole layer. The distances between the top/bottom gates and the electron/hole layers are set to be equal to $d_g$. The voltages applied to the two layers $V_e$, $V_{h}$ are used to tune the overall chemical potential and the exciton chemical potential.
  (b) The exciton chemical potential $\mu_{ex}=eV_b-E_g$ is determined by the inter-layer bias voltage $V_b=V_h-V_e$ while the overall chemical potential $\mu$ is tuned by $(V_e+V_h)/2$ to make the system charge neutral.
  (c) When a magnetic field is applied in the $z$ direction, the non-interacting electron and hole bands are quantized to LLs. At CNP, the overall chemical potential must lay between the electron and hole LLs with the same index, for example, the $N$-th level. If the electron-hole interaction is considered, particle-hole excitations will bind to form exciton with binding energy $E_B$. If the gap between the $N$-th electron and hole LLs is smaller than $E_B$, such exciton of LLs will spontaneously form and condense.
  }
  \label{fig:set_up}
\end{figure}

Two-dimensional bilayer separated by a perfect insulating barrier is expected to be a candidate system to realize exciton condensation at charge neutrality point (CNP) where the two layers are equally charged by electrons and holes\cite{butovCondensationIndirectExcitons1994,zhuExcitonCondensateSemiconductor1995,littlewoodPossibilitiesExcitonCondensation1996,butovExcitonCondensationCoupled2003,littlewoodModelsCoherentExciton2004,minRoomtemperatureSuperfluidityGraphene2008,foglerHightemperatureSuperfluidityIndirect2014a,wuTheoryTwodimensionalSpatially2015,highSpontaneousCoherenceCold2012,liExcitonicSuperfluidPhase2017,duEvidenceTopologicalExcitonic2017,wangElectricalTuningInterlayer2018,xieElectricalReservoirsBilayer2018}.
This excitonic insulator (EI) phase was realized recently in the dual-gated transition metal dichalcogenide (TMD) double layers\cite{wangEvidenceHightemperatureExciton2019a,maStronglyCorrelatedExcitonic2021}.
The experimental setup is illustrated in Fig. \ref{fig:set_up}(a), where the electron layer (blue) and hole layer (orange) are sandwiched between the top and bottom gates (black), and dielectric spacers (gray) are inserted between gates and layers to avoid direct tunneling.
The gate-layer voltage $(V_{e}+V_{h})/2$ is used to control the overall chemical potential $\mu$ to make the system charge neutral.
And the exciton density (charge number density per layer) $n_{ex}$ is tuned by exciton chemical potential $\mu_{ex}=eV_b-E_{g}$ where $V_{b}=V_h-V_e$ is the inter-layer bias voltage and $E_g$ is the spatially indirect gap between the electron and hole bands at zero bias.

Low energy excitations of single-layer TMD near the valley center are approximated as free fermions with quadratic dispersion.
By tuning exciton chemical potential $\mu_{ex}$, a typical non-interacting band structure at CNP is illustrated in Fig. \ref{fig:set_up}(b), where the electron and hole layers have nested Fermi surfaces.
In the absence of single-particle tunneling $t$, the electron and hole layers have charge conservation separately and the system has a $\mathrm{U}_{e}(1)\times \mathrm{U}_{h}(1)$ symmetry.
However, when inter-layer excitons are generated and condensed due to the attractive interaction between electrons and holes, a non-zero mean-field inter-layer coherence $\Delta \equiv h^{mf}_{eh} =|\Delta|\re^{i\phi}$ will spontaneously arise, break the electron-hole $\mathrm{U}(1)$ symmetry and leave only the total charge conservation.
Besides, the inter-layer coherence $\Delta$ will also gap out the Fermi surfaces and drive the system into an excitonic insulator phase\cite{Jerome1967}.
Due to the spontaneous symmetry breaking, the long wave phase fluctuation of the excitonic order parameter in real space $\delta\phi(\bm{r})$ is the Goldstone mode and related to the exciton superfluidity.
In real materials, a tiny single-particle tunneling $t$ is unavoidable which breaks the electron-hole $\mathrm{U}(1)$ symmetry initially
This will pin the phase of the inter-layer coherence to $\phi=\arg t$, gap out the zero energy Goldstone mode, and destroy the exciton superfluidity\cite{eisensteinBoseEinsteinCondensation2004,nandiExcitonCondensationPerfect2012,zhuGateTuningExciton2019}.
Without a dielectric spacer, the single-particle tunneling strength in TMDs bilayer is in the order of $10\mathrm{meV}$\cite{tongTopologicalMosaicsMoire2017,zhuGateTuningExciton2019}.
By inserting a few layer hBN spacer between the two TMD single layers, the inter-layer hopping strength will be exponentially suppressed.
Since the single-particle tunneling is unavoidable (although could be very small), the electron-hole bilayer could be considered as an excitonic insulator only when $|\Delta|\gg t$ is satisfied.
In addition to the phase pinning effect, the inter-layer tunneling will also induce a tunneling current when the circuit is closed, which drives the system into a non-equilibrium state.
However, as long as $t$ is small enough, the tunneling current is insignificant and the non-equilibrium transport physics could be ignored.

When magnetic field is applied along the $z$ direction, the parabolic dispersions of electron and holes are quantized into Landau levels (LLs).
At CNP, the overall chemical potential must lay between the electron and hole LLs with the same index as illustrated in Fig. \ref{fig:set_up}(c).
The low-energy excitations are free particle-hole pairs between the highest occupied electron LL and the highest empty hole LL. 
When interaction is considered, such free pairs will bind to form exciton of LLs with binding energy $E_B$.
By tuning magnetic field $B$ or exciton chemical potential $\mu_{ex}$ to make the exciton binding energy $E_B$ larger than the gap between the highest occupied electron and empty hole LLs, excitons of LLs will spontaneously form and condense.
Since the gap between the highest electron and hole LLs will oscillate with $1/B$ and $\mu_{ex}$, physical properties of the exciton condensation state will also oscillate.
As an insulator, the conventional quantum oscillation of resistance might be hard to detect.
In our paper, we will focus on the inter-layer capacitance 
\begin{equation}
  C_{I}=e^2\left(\frac{\partial n_{ex}}{\partial \mu_{ex}}\right)_T\label{eq:cap_inter_layer}
\end{equation}
to show the quantum oscillation phenomenon in such excitonic insulating electron-hole bilayer system.

There are several advantages of the inter-layer capacitance measurement.
Firstly, it's unique to the bilayer system and could be measured accurately in real experiments\cite{maStronglyCorrelatedExcitonic2021}.
Besides, as we will show, the oscillation behaviors of the inter-layer capacitance could help us to distinguish an excitonic gap from a single-particle one.
When the magnetic field is so large that the cyclotron energy $\hbar(\omega_e+\omega_h)$ is much larger than the exciton binding energy, one can always tune the exciton chemical potential $\mu_{ex}$ to make $E_B$ smaller than the LL direct gap and exciton will not spontaneously generate and condense anymore.
For such a situation, the bilayer system in the magnetic field is just two independent quantum Hall liquids and is charge incompressible at zero temperature\cite{zeeQuantumHallFluids1995} which results in a zero inter-layer capacitance $C_{I}(T=0)=0$.
In other words, the inter-layer coherence $\Delta$ of an excitonic insulator could be destroyed by a strong magnetic field and the inter-layer capacitance might oscillate to zero.
While for a consistent hybridization from single-particle tunneling, the inter-layer capacitance will never be zero.

\section{Model And Mean-field Theory}
Without magnetic field, the many-body Hamiltonian for the bilayer system as illustrated in Fig. \ref{fig:set_up}(a) is modeled as\cite{wuTheoryTwodimensionalSpatially2015,zengElectricallyControlledTwodimensional2020a} 
\begin{subequations}
  \begin{align}
    H_0=&\sum_{ss'=eh,\bm{k}}(h^0_{ss'\bm{k}}-\mu\delta_{ss'})c^{\dagger}_{s\bm{k}}c_{s'\bm{k}},\\
    H_{I}=&\frac{1}{2\mathcal{V}}\sum_{ss'=eh}\sum_{\bm{k}_1\bm{k}_2\bm{q}}V_{ss'}(\bm{q})c^{\dagger}_{s\bm{k}_1}c^{\dagger}_{s'\bm{k}_2}c_{s'\bm{k}_2+\bm{q}}c_{s\bm{k}_1-\bm{q}},\label{eq:interaction_plane_wave}
  \end{align} \label{eq:many-body}
\end{subequations}
where $c^{\dagger}_{e\bm{k}}$ and $c^{\dagger}_{h\bm{k}}$ are electron creation operators in the electron and hole layer, $\mathcal{V}\equiv L_xL_y$ is the area of the 2D system and $L_i$ is the system length in the $i$ direction.
Under $k\cdot p$ approximation, the single-particle Hamiltonian is
\begin{equation}
  h^0_{\bm{k}}=\begin{bmatrix}
    \hbar^2k^2/2m_e-\mu_{ex} & t \\
    t^* & -\hbar^2k^2/2m_h
  \end{bmatrix},\label{eq:single-particle}
\end{equation}
where $m_{e/h}$ are the effective masses and $t$ is the inter-layer tunneling strength.
The intra- and inter-layer interactions are taken as the gate-screened Coulomb interaction\cite{hallamScreeningElectronelectronInteraction1996} $V(q)\equiv V_{s=s'}(q)\approx 2\pi e^2/\epsilon q (1-\re^{-2\kappa q d_{g}})$ and $U(q)\equiv V_{s\ne s'}(q)\approx V(q)\re^{-\kappa q d}$ where $\epsilon=\sqrt{\epsilon^{xy}\epsilon^{z}}$ is the effective dielectric constant and $\kappa\equiv \sqrt{\epsilon^{xy}/\epsilon^{z}}$ is the anisotropy parameter (a detailed derivation could be found in Appendix \ref{app:gate-screen}).

By assuming a non-zero EI order parameter $\rho_{eh\bm{k}}$ where $\rho_{ss'\bm{k}}\equiv\ave{c^{\dagger}_{s'\bm{k}}c_{s\bm{k}}}-\delta_{ss'}\delta_{sh}$ is the density matrix relative to the uncharged state ($\rho_0=\delta_{ss'}\delta_{sh}$ is subtracted to avoid double counting\cite{shimSpinorbitInteractionsBilayer2009,wuTheoryTwodimensionalSpatially2015}), the interacting part of the many-body Hamiltonian Eq. \eqref{eq:many-body} is decoupled into a non-interacting mean-field Hamiltonian
\begin{equation}
  H_{MF}=\sum_{ss'\bm{k}}(h^0_{ss'\bm{k}}+h^H_{ss'}+h^F_{ss'\bm{k}}-\mu\delta_{ss'})c^{\dagger}_{s\bm{k}}c_{s'\bm{k}}.
\end{equation}
The Hartree and Fock terms are constructed by density matrix as
\begin{subequations}
  \begin{gather}
    h^H=\frac{e^2n_{ex}}{2C_{geo}}\sigma_z, \label{eq:single_Hartree}\\
    h^F_{ss'\bm{k}}=-\frac{1}{\mathcal{V}}\sum_{\bm{k}}V_{ss'}(\bm{k}-\bm{k}')\rho_{ss'\bm{k}'},
  \end{gather}\label{eq:single_HF}
\end{subequations}
where $\sigma_z$ is the Pauli matrix $n_{ex}=\mathcal{V}^{-1}\sum_{\bm{k}}\rho_{ee\bm{k}}$ is exciton density and $C_{geo}=\epsilon^z/4\pi d$ is the geometry capacitance of the charged electron-hole double layer.
The mean-field Hamiltonian $h^{MF}_{\bm{k}}=h^0_{\bm{k}}+h^H+h^F_{\bm{k}}$ is a $2\times 2$ matrix and has two eigenvalues, i.e.
\begin{equation}
  h^{MF}_{\bm{k}}\ket{c/v,\bm{k}}=\xi_{c/v,\bm{k}}\ket{c/v,\bm{k}},\label{eq:diagonalization}
\end{equation}
where $\xi_{c,\bm{k}}>\xi_{v,\bm{k}}$ are the mean-field energy bands and $\ket{c/v,\bm{k}}$ are the corresponding eigenstates.
Then the new density matrix could be reconstructed as
\begin{equation}
  \rho_{\bm{k}}=\sum_{i=c,v}f_{i,\bm{k}}(\mu)\ket{i,\bm{k}}\bra{i,\bm{k}}-\delta_{ss'}\delta_{sh},\label{eq:new_dst_mat}
\end{equation}
where $f_{i,\bm{k}}(\mu)=1/[1+\re^{(\xi_{i,\bm{k}}-\mu)/k_BT}]$ are the occupation numbers.
By requiring charge neutrality, the overall chemical potential is determined by solving 
\begin{equation}
  \sum_{\bm{k}}[f_{c,\bm{k}}(\mu)+f_{v,\bm{k}}(\mu)-1]=0.\label{eq:chemical_potential_determine}
\end{equation}
Eq. \eqref{eq:single_HF}\eqref{eq:diagonalization}\eqref{eq:new_dst_mat}\eqref{eq:chemical_potential_determine} form the full self-consistent procedure.
At zero temperature $T=0$, Eq. \eqref{eq:chemical_potential_determine} is simply solved as $f_{c,\bm{k}}=0$ and $f_{v,\bm{k}}=1$.

When a magnetic field is applied along the $z$ direction, it's more convenient to adopt the LL basis. 
In Landau gauge $\bm{A}=(-yB,0)$, the parabolic bands quantized into LLs $\ket{\phi_{nk_x}}$ as shown in Fig. \ref{fig:set_up}(c), where $n$ is the LL index and $k_x$ is the momentum in $x$ direction.
By defining the creation operators for LL electrons $l^{\dagger}_{snk_x}\equiv \sum_{\bm{k}'}\inp{\bm{k}'}{\phi_{nk_x}}c^{\dagger}_{s\bm{k}'}$ which in fact is a basis transformation, the many-body Hamiltonian with magnetic field is written under the LL basis as 
\begin{subequations}
  \begin{align}
    H_0=&\sum_{ss'nk_x}(h^0_{n,ss'}-\mu\delta_{ss'})l^{\dagger}_{snk_x}l_{s'nk_x},\\
    H_I=&\frac{1}{2\mathcal{V}}\sum_{ss'n_ik_i}\sum_{\bm{q}}V_{ss'}(\bm{q})\re^{iq_y(k_1-k_2)\ell^2}\Lambda_{n_4n_1}^*(\bm{q})\Lambda_{n_2n_3}(\bm{q})\nonumber\\
    &\times l^{\dagger}_{sn_1k_1+q_x/2}l^{\dagger}_{s'n_2k_2-q_x/2}l_{s'n_3k_2+q_x/2}l_{sn_4k_1-q_x/2},
  \end{align}\label{eq:manybody_landau_level}
\end{subequations}
where $\ell=\sqrt{\hbar/eB}$ is the magnetic length and $\Lambda_{mn}(\bm{q})$ is the form factor of LLs defined by\cite{macdonaldInfluenceLandaulevelMixing1984}
\begin{equation}
  \Lambda_{mn}(\bm{q})\equiv\bra{\phi_{mk-q_x/2}}\re^{-i\bm{q}\cdot{\bm{r}}}\ket{\phi_{nk+q_x/2}}\re^{ikq_y\ell^2}.
\end{equation}
The single-particle Hamiltonian now becomes
\begin{equation}
  h^0_{n}=\begin{bmatrix}
    \hbar\omega_e(n+1/2)-\mu_{ex} & t\\ 
    t^* & -\hbar\omega_h(n+1/2)
  \end{bmatrix},\label{eq:single-magnetic}
\end{equation}
where $\omega_s\equiv eB/m_s$ are the cyclotron frequency.
Details of deriving Eq. \eqref{eq:manybody_landau_level} are given in Appendix \ref{app:many-body_landau_level}.

The density matrix is now defined as $\rho_{sn_1,s'n_2}(k_x)\equiv\ave{l^{\dagger}_{s'n_2k_x}l_{sn_1k_x}}-\delta_{ss'}\delta_{sh}$.
However, due to symmetry constraints, not all the elements survive.
Although the vector potential in Landau gauge $\bm{A}=(-yB,0)$ breaks translation symmetry in $y$ direction, the physics is expected to be independent of the choice of gauge.
After a small translation in $y$ direction, i.e. $\bm{A}\to (-(y-\eta)B,0)$, the magnetic field is invariant while the LL electron transforms as $l^{\dagger}_{snk_x}\to l^{\dagger}_{snk_x+eB\eta/\hbar}$.
It's easy to see that the many-body Hamiltonian Eq. \eqref{eq:manybody_landau_level} is invariant under such magnetic translation while the density matrix transforms from $\rho_{sn_1,s'n_2}(k_x)$ to $\rho_{sn_1,s'n_2}(k_x+eB\eta/\hbar)$.
By requiring magnetic translation symmetry in $y$ direction, the density matrix should be $k_x$-independent, i.e., $\rho_{sn_1,s'n_2}(k_x)=\rho_{sn_1,s'n_2}$.
As discussed in Appendix \ref{app:mean_field_channels}, when magnetic translation symmetry is preserved, the EI order parameters $\rho_{en_1,hn_2}$ could be decomposed into independent channels labeled by its angular momentum $M\equiv n_1-n_2$.
In the charge neutral case, the overall chemical potential $\mu$ must lay between electron and hole LLs with the same index, for example, the $N$-th level as illustrated in Fig. \ref{fig:set_up}(c).
At this time, the $s$-wave pairing case with zero angular momentum $M=0$ usually has the lowest energy.
For electron and hole bands with trivial band topology, high angular momentum exciton condensation in the quantum Hall regime is energetically preferable only when the electron and hole layers are charge imbalanced as investigated by \citet{zouElectricalControlTwoDimensional2023}.
In summary, by requiring magnetic translation symmetry and $s$-wave pairing, the only surviving density matrix elements are $\rho_{sn,s'n}$ and they will be abbreviated as $\rho_{n,ss'}$ in the following text.

Once the mean-field channels are determined, the Hartree Fock procedure is straightforward and the mean-field Hamiltonian in LL basis is written as
\begin{equation}
  H_{MF}=\sum_{ss'nk_x}(h^0_{n,ss'}+h^H_{ss'}+h^F_{n,ss'}-\mu\delta_{ss'})l^{\dagger}_{snk_x}l_{s'nk_x}.
\end{equation}
Since the Hartree term is just a renormalization of the exciton chemical potential due to the geometry electrostatic energy, it's independent of basis transformation and is still given by Eq. \eqref{eq:single_Hartree}.
The only difference is that the exciton density is calculated as $n_{ex}={(2\pi \ell^2)^{-1}}\sum_n \rho_{n,ee}$.
The Fock term becomes 
\begin{equation}
  h^F_{n,ss'}=-\sum_{n'}V_{ss',nn'}\rho_{n',ss'},
\end{equation}
where $V_{ss',nn'}=\mathcal{V}^{-1}\sum_{\bm{q}}V_{ss'}(q)|\Lambda_{n'n}(\bm{q})|^2$ is the interaction matrix elements projected to LL basis.
By replacing the $\bm{k}$ index in \eqref{eq:diagonalization}\eqref{eq:new_dst_mat}\eqref{eq:chemical_potential_determine} with LL index $n$, we get the full self-consistent equations under LL basis.

\section{Results}
In our calculation, the parameters are set to be consistent with the MoSe$_2$/hBN/WSe$_2$ heterostructure experimentally studied by \citet{maStronglyCorrelatedExcitonic2021}.
The effective masses of the conduction band minimum of MoSe$_2$ and valence band maximum of WSe$_2$ at the $K$-valley centers are about $m_e\approx 0.58 m_0$, $m_h\approx 0.36m_0$\cite{kormanyosTheoryTwodimensionalTransition2015a} ($m_0$ is the bare electron mass).
The inter-layer and gate-layer distances are taken as $d\approx 2.5\mathrm{nm}$ ($5\sim 6$ hBN spacer) and $d_g\approx 10\mathrm{nm}$.
The dielectric constant of hBN is about $\epsilon^{xy}\approx 6.71$ and $\epsilon^{z}\approx 3.57$\cite{caiInfraredReflectanceSpectrum2007}.
Thus the anisotropy parameter and the effective dielectric constant are about $\kappa\approx 1.37$, $\epsilon\approx 4.89$.
To fit the inter-layer exciton binding energy in the experiment\cite{maStronglyCorrelatedExcitonic2021} (about 20meV), a larger effective dielectric constant $\epsilon=9$ is used in the calculation.

\subsection{Inter-layer Capacitance at Zero Magnetic Field}

\begin{figure}[h]
  \centering
  \includegraphics[width=\linewidth]{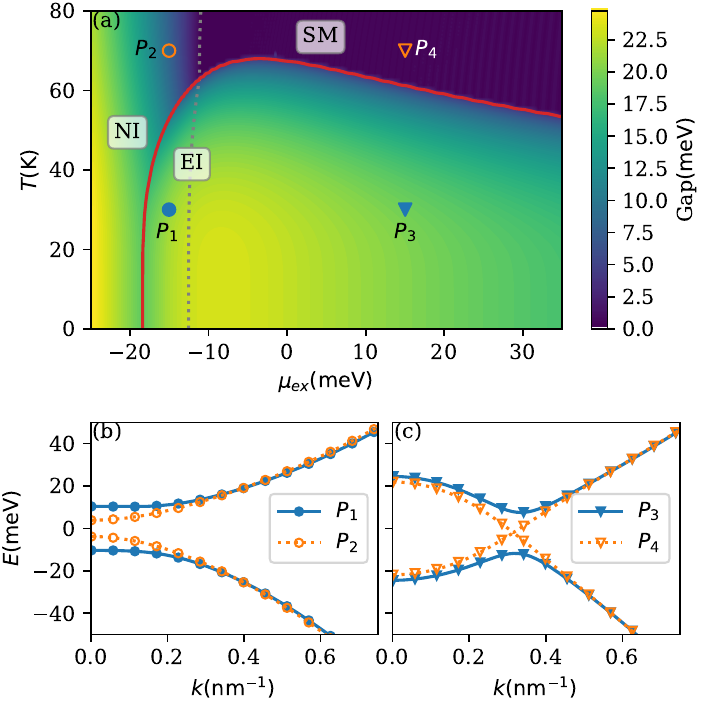}
  \caption{(a) Zero magnetic field phase diagram as a function of exciton chemical potential $\mu_{ex}$ and temperature $T$, the single-particle tunneling strength $t$ is assumed to be zero.
    Below the red solid line, the EI order parameter is not zero, i.e. $\rho_{eh}\ne 0$ which means the system is in the EI phase while above the red solid line $\rho_{eh}= 0$ and the system is in the normal phase.
    The gray dashed line further separates the normal phase into a NI phase and a SM phase.
    In the NI phase, there is no inversion between the renormalized electron and hole bands, while in the SM phase electron and hole bands are inverted.
    The color represents the mean-field band gap.
    (b)(c) Typical mean-field band structures in different regions of the parameter space, points $P_{1-4}$ in (a) are used for example.
    }
  \label{fig:phase_T_mu}
\end{figure}

Let's first ignore the single-particle tunneling $t$.
At zero magnetic field, the mean-field phase diagram as a function of exciton chemical potential $\mu_{ex}$ and temperature $T$ is calculated and plotted in Fig. \ref{fig:phase_T_mu}(a).
The red solid line is the boundary of the region $\rho_{eh}(T,\mu_{ex})\ne 0$.
The area below the red line is the EI phase with a non-zero order parameter $\rho_{eh}\ne 0$.
While above the red line, there is no EI order and the system is in the normal phase.
The gray dashed line is determined by requiring the renormalized offset between the electron and hole bands to be equal to the original gap, after which the inversion between the renormalized conduction and valence bands from different layers occurs.
To the left of the gray dashed line, there is no band inversion and the normal phase is just a normal insulator (NI) and to the right of this line, the normal phase is a semi-metal (SM).
In the EI phase, the gray dashed line does not mark a phase transition but rather indicates a BEC-BCS crossover to some extent.
By diagonalizing the mean-field Hamiltonian $h_{\bm{k}}^{MF}$, mean-field band structures are obtained and the gap is represented by the color plot in Fig. \ref{fig:phase_T_mu}(a).
Besides, typical mean-field band structures in different regions of the parameter space are also plotted in Fig. \ref{fig:phase_T_mu}(b)(c) (points $P_{1-4}$ in Fig. \ref{fig:phase_T_mu}(a) are used for example).

\begin{figure}[h]
  \centering
  \includegraphics[width=\linewidth]{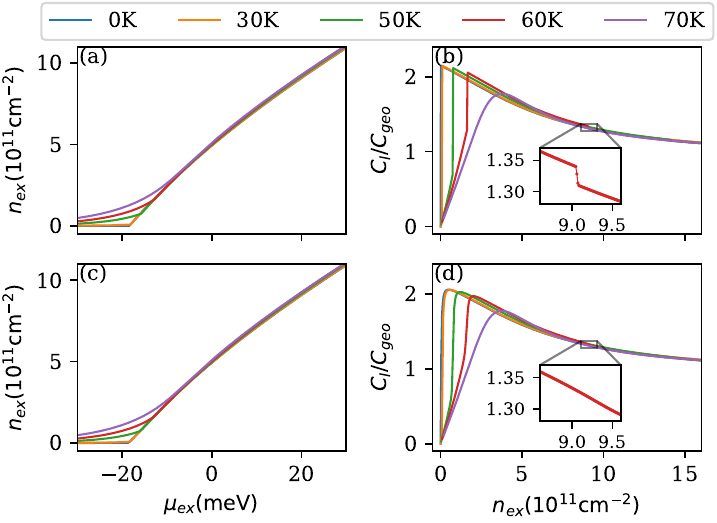}
  \caption{(a) Exciton density as a function of exciton chemical potential and temperature at CNP.
  The exciton chemical potential $\mu_{ex}$ is used as abscissa and different temperatures are represented by different color lines.
  (b) Inter-layer capacitance calculated by Eq. \eqref{eq:cap_inter_layer} as a function of exciton density and temperature.
  The inset shows a magnified view of the line at 60K near the phase boundary between the EI and SM phases.
  (c)(d) The same quantities as in (a)(b) except that a finite single-particle tunneling strength $t\approx 0.01\mathrm{meV}$ instead $t=0\mathrm{meV}$ is used.
  }
  \label{fig:cap_T_mu}
\end{figure}

Then the exciton density at CNP is calculated and plotted as a function of exciton chemical potential $\mu_{ex}$ (the abscissa) and temperature (different color lines) in Fig. \ref{fig:cap_T_mu}(a).
Using the definition Eq. \eqref{eq:cap_inter_layer}, the inter-layer capacitance $C_{I}$ (in the unit of geometry capacitance $C_{geo}=\epsilon^{z}/4\pi d$) is also calculated and plotted in Fig. \ref{fig:cap_T_mu}(b) as a function of exciton density.
The inset in Fig. \ref{fig:cap_T_mu}(b) shows a magnified view of the line near the phase boundary between the EI phase and the SM phase at 60K.
Due to the exchange part of the interaction which accounts for exciton condensation, the inter-layer capacitance is greatly enhanced from its classic geometry value, which is consistent with previous studies\cite{wuTheoryTwodimensionalSpatially2015,zengElectricallyControlledTwodimensional2020a}.
Besides, discontinuities of $C_{I}$ are shown at the transition points between EI and normal phases.
However, these discontinuities may be absent in real experiments.
On the one hand, the transition between EI and NI in the low-density region at finite temperature is a BKT transition\cite{foglerHightemperatureSuperfluidityIndirect2014a,filinovBerezinskiiKosterlitzThoulessTransitionTwoDimensional2010} that is beyond the mean-field description, and its main effect is to smooth out the dramatic changes in the mean-field theory.
On the other hand, these discontinuities are easily smoothed by a very small single-particle tunneling effect.
In Fig. \ref{fig:cap_T_mu}(c)(d), the same quantities as in Fig. \ref{fig:cap_T_mu}(a)(b) are plotted, except that a finite single-particle tunneling strength $t\approx 0.01\mathrm{meV}$ is used.
Although the tunneling strength $t$ is much smaller than the mean-field gap (about $20\mathrm{meV}$ as indicated in Fig. \ref{fig:phase_T_mu}(a)), the discontinuities at the EI phase boundary no longer exist as shown in Fig. \ref{fig:cap_T_mu}(d).

\subsection{Quantum Oscillation of the Inter-layer Capacitance}\label{sec:oscillation}

\begin{figure}[h]
  \centering
  \includegraphics[width=\linewidth]{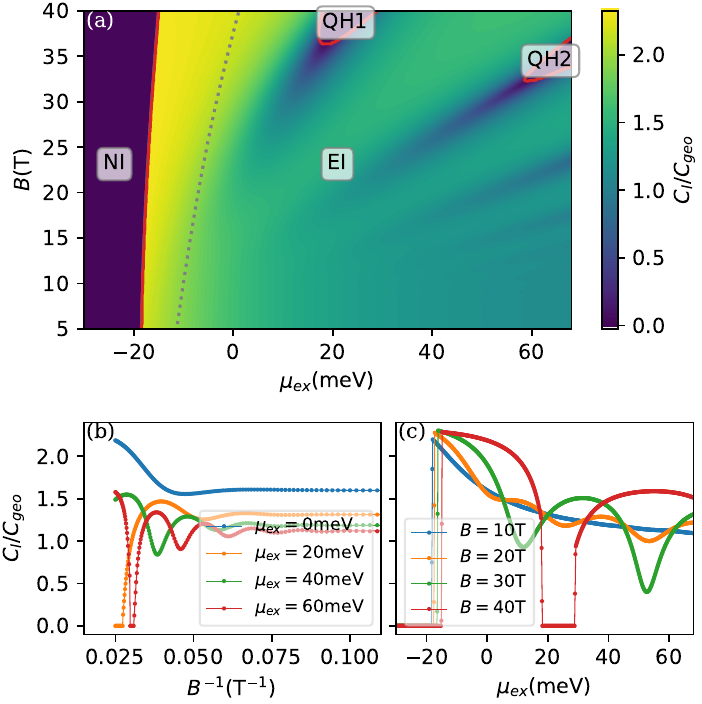}
  \caption{(a) Phase diagram as a function of exciton chemical potential $\mu_{ex}$ and magnetic field strength $B$ at 0K. 
  The single-particle tunneling strength is assumed to be zero.
  The red solid line separates the region into EI phase with a nonzero EI order parameter $\rho_{eh}\ne 0$ and normal phases where $\rho_{eh}= 0$.
  The gray dashed line is the critical line for band inversion.
  In the NI phase, all hole LLs are occupied and all electron LLs are empty.
  In the QH-$N$ phases, the first $N$ electron LLs are occupied while the first $N$ hole LLs are empty.
  The pseudo-color map represents the inter-layer capacitance $C_{I}/C_{geo}$.
  (b)(c) Oscillations of inter-layer capacitance $C_{I}$ versus $B^{-1}$ and $\mu_{ex}$ respectively.}
  \label{fig:phase_B_mu}
\end{figure}

Since the inter-layer tunneling gap out the Goldstone mode, the BKT phenomenon is suppressed for temperatures much below the energy scale of the tunneling strength $t$, especially at zero temperature.
In this situation, the mean-field theory is still qualitatively right.
Thus in this part, we will focus on zero temperature.
Ignoring the single-particle tunneling effect, the mean-field phase diagram as a function of exciton chemical potential $\mu_{ex}$ and magnetic field strength $B$ is plotted in Fig. \ref{fig:phase_B_mu}(a).
Similar to before, the red solid line is the boundary of the region $\rho_{eh}\ne 0$ and the gray dashed line is the critical line for band inversion.
Only in the EI phase, $\rho_{eh}\ne 0$ and there is an inter-layer coherence.
In the NI phase, there is no band inversion between the electron and hole bands where all the hole LLs are occupied and the electron LLs are empty.
In the QH phase, according to the index $N$ of the highest inverted electron and hole LLs, the regions in the parameter space are labeled by QH-$N$ as shown in Fig. \ref{fig:phase_B_mu}(a).
The color in Fig. \ref{fig:phase_B_mu}(a) represents the inter-layer capacitance $C_{I}/C_{geo}$, which is plotted in more detail in Fig. \ref{fig:phase_B_mu}(b)(c).
Oscillations versus $B^{-1}$ and $\mu_{ex}$ are easily identified.
Similar to the quantum oscillation in metal, the oscillation frequency versus $B^{-1}$ increases with exciton chemical potential as shown in Fig. \ref{fig:phase_B_mu}(b).
It is also worth noting that inter-layer capacitance oscillates to zero in the QH phases, which reflects the fact that a QH state is charge incompressible at zero temperature.

\begin{figure}[h]
  \centering
  \includegraphics[width=\linewidth]{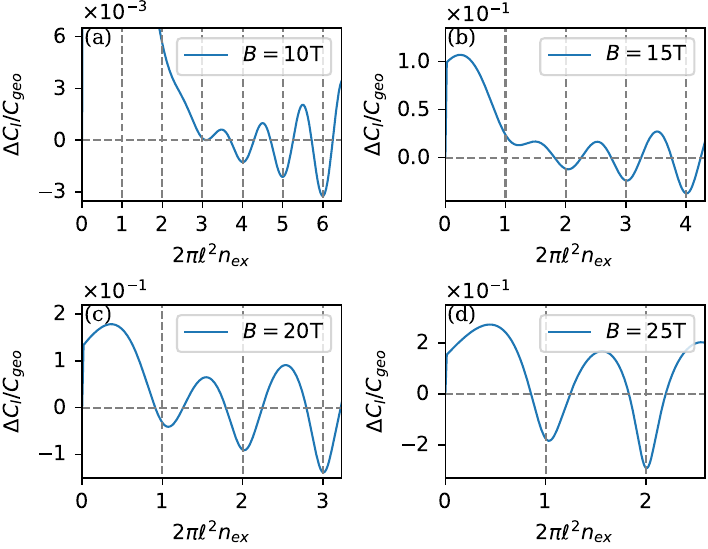}
  \caption{Oscillations of $\Delta C_{I}(n_{ex},B)$ versus $n_{ex}$ for different magnetic field strengths.}
  \label{fig:cap_oscillation}
\end{figure}

To see the oscillations versus $\mu_{ex}$ more clearly, let's transform the abscissa from $\mu_{ex}$ to $n_{ex}$ and define $\Delta C_{I}(n_{ex},B)$ as
\begin{equation}
  \Delta C_{I}(n_{ex},B)\equiv C_I(n_{ex},B)-C_I(n_{ex},B=0).
\end{equation}
Then the oscillations of $\Delta C_{I}$ versus $n_{ex}$ are shown in Fig. \ref{fig:cap_oscillation}(a-d) for different magnetic field strengths.
A period about $(2\pi\ell^2)^{-1}=eB/h$ is observed which is exactly the LL degeneracy for a spinless fermion.

\section{Summary and Discussion}
For an electron-hole bilayer without any inter-layer coupling, the system is just a semi-metal, and quantum oscillations are not surprising due to the Landau quantization of the electron and hole Fermi surfaces\cite{shoenbergMagneticOscillationsMetals2009}.
An inter-band hybridization $h_{eh}$ will gap out the Fermi surfaces and lead the system into an insulating phase at CNP.
However, as long as the hybridization strength is comparable with the cyclotron frequency $\hbar (\omega_e+\omega_h)$, quantum oscillations of physical quantities are still expected.
Such oscillations have already been predicted\cite{knolleQuantumOscillationsFermi2015,zhangQuantumOscillationNarrowGap2016,ongQuantumOscillationsInsulator2018,xiangQuantumOscillationsElectrical2018,pandaQuantumOscillationsMagnetization2022} and detected\cite{suzukiLandauLevelHybridizationQuantum2004,tanUnconventionalFermiSurface2015,hanAnomalousConductanceOscillations2019,xiaoAnomalousQuantumOscillations2019} in narrow gap insulators where the hybridization has a single-particle origination.
While in this paper, we show that quantum oscillations will also appear in EI systems where the inter-band hybridization purely arises from exciton condensation. 

A more interesting observation is the QH phases in the phase diagram Fig. \ref{fig:phase_B_mu}(a) where there is no EI order parameter.
And these phases are also noted in a similar study by \citet{zouElectricalControlTwoDimensional2023}.
From the physical picture ``exciton formed by electron/hole LLs'' illustrated in Fig. \ref{fig:set_up}(c), the critical magnetic field strength could be estimated by requiring the cyclotron frequency to be comparable to the exciton binding energy, i.e. $\hbar(\omega_e+\omega_h)/2=E_B$, which implies
\begin{equation}
  B_c=\frac{2m_em_h}{m_e+m_h}\frac{E_B}{e\hbar}\label{eq:criticle_magnetic}.
\end{equation}
Substitute the parameters $m_e=0.58m_0$, $m_h=0.36m_0$ and value of the zero density binding energy $E_B\approx 20\mathrm{meV}$ to Eq. \eqref{eq:criticle_magnetic}, and the critical field strength is estimated as $B_c\approx 77\mathrm{T}$.
We argue that this value is an overestimation since the binding energy usually drops with the increase of exciton density $n_{ex}$ (or equivalently, exciton chemical potential $\mu_{ex}$)\cite{zengElectricallyControlledTwodimensional2020a}.
This point can also be seen from the fact that the critical field strength of the QH2 phase is lower than the QH1 phase.

The emergence of the QH phases also reflects the instability of an excitonic gap, which could be destroyed not only by temperature\cite{Jerome1967} and electrical field\cite{shaoElectricalBreakdownExcitonic2023} but also by magnetic field.
Such instability is a key difference between exciton gap and single-particle gap and could be easily identified in inter-layer capacitance measurements.
As shown in Fig. \ref{fig:phase_B_mu}(b)(c), the inter-layer capacitance at zero temperature oscillates to zero when the EI order and inter-band hybridization are destroyed by the magnetic field in the QH phases.
While the capacitance will never be zero if the gap has a single-particle origination.
To see this, let's assume a non-zero single-particle tunneling strength $t\ne 0$ and ignore the exchange part of the interaction which accounts for the exciton condensation.
Then the charge density per-layer is calculated as 
\begin{equation}
  n_{ex}=\frac{1}{4\pi\ell^2}\sum_{n}\left[1-\frac{\hbar\omega^*(2n+1)-\tilde{\mu}_{ex}}{\sqrt{[\hbar\omega^*(2n+1)-\tilde{\mu}_{ex}]^2+4t^2}} \right],
\end{equation}
where $\omega^*=(\omega_e+\omega_h)/2$ and $\tilde{\mu}_{ex}=\mu_{ex}-e^2n_{ex}/C_{geo}$ is the renormalized ``exciton chemical potential'' by inter-layer geometry electrostatic energy.
By definition, the inter-layer capacitance should be calculated as 
\begin{equation}
  C_{I}\equiv e^2\frac{\partial n_{ex}}{\partial {\mu_{ex}}}=e^2\frac{\partial n_{ex}}{\partial{\tilde{\mu}_{ex}}}\frac{\partial\tilde{\mu}_{ex}}{\partial \mu_{ex}}=e^2\frac{\partial n_{ex}}{\partial {\tilde{\mu}_{ex}}}\left(1-\frac{C_{I}}{C_{geo}}\right).
\end{equation}
Denote $\tilde{C}_{I}=e^2\partial n_{ex}/\partial \tilde{\mu}_{ex}$, and the inter-layer capacitance is solved as $C_{I}=(\tilde{C}_{I}^{-1}+C^{-1}_{geo})^{-1}$.
It's easily verified that 
\begin{equation}
  \tilde{C}_{I}=\frac{e^2}{4\pi \ell^2}\sum_{n}\frac{4t^2}{\{[\hbar\omega^*(2n+1)-\tilde{\mu}_{ex}]^2+4t^2\}^{3/2}}>0
\end{equation}
as long as the hybridization strength $t$ is nonzero, and the inter-layer capacitance must satisfy $0<C_{I}<C_{geo}$.
That's to say, if the inter-layer hybridization has a single-particle origination, the inter-layer capacitance will never oscillate to zero at zero temperature.
So a zero inter-layer capacitance in a strong magnetic field could be used to exclude the single-particle contribution to the excitonic gap, which is an essential requirement for exciton superfluidity.

\begin{acknowledgments}

  We thank Prof. Kin Fai Mak for helpful discussions.
  This work was fully supported by a fellowship award from the Research Grants Council of the Hong Kong Special Administrative Region, China (Project No. HKUST SRFS2324-6S01)

\end{acknowledgments}

\appendix

\section{Gate Screening Interaction}\label{app:gate-screen}
When the gate layer distance $d_g$ is comparable to the inter-layer distance $d$, the screening effects from the gates are not negligible.
To derive the gate screening interaction, let's solve the Possion equation of a point charge.
For convenience, let's assume the point charge is in the electron layer, using the Dirichlet boundary condition, the Possion equation reads
\begin{subequations}
  \begin{gather*}
    \epsilon^{xy}\nabla^2_{\bm{r}}\varphi(\bm{r},z)+\epsilon^z\partial_z^2\varphi(\bm{r},z)=-4\pi e\delta(\bm{r}-\bm{r}_0)\delta(z-d-d_g),\\
    \nabla_{\bm{r}}\varphi(\bm{r},z)\big|_{z=0,d+2d_g}=\bm{0}.
  \end{gather*}
\end{subequations}
Define the 2D Fourier transformation of $\varphi(\bm{r},z)$ as $\varphi_{\bm{q}}(z)=\int\rd\bm{r}\;\varphi(\bm{r},z)\re^{-i\bm{q}\cdot\bm{r}}$, the Possion equation becomes 
\begin{subequations}
  \begin{gather}
    \epsilon^z\partial_z^2\varphi_{\bm{q}}(z)-q^2\epsilon^{xy}\varphi_{\bm{q}}(z)=-4\pi e\delta(z-d-d_g),\\
    \varphi_{\bm{q}}(z=0,d+2d_g)=0.
  \end{gather}\label{eq:possion_eq}
\end{subequations}
Define the effective dielectric constant and anisotropy parameter as $\epsilon=\sqrt{\epsilon^{xy}\epsilon^z}$, $\kappa=\sqrt{\epsilon^{xy}/\epsilon^z}$, and the Possion equation Eq. \eqref{eq:possion_eq} is solved as 
\begin{gather*}
  \varphi_{\bm{q}}(z)=\frac{2\pi}{\epsilon q}[c_1\re^{\kappa q z}+c_2\re^{-\kappa q z}+\re^{-\kappa q|z-(d+d_g)|}],\\
  c_1=-\frac{\re^{-\kappa q(2d_g+d)}\sinh{\kappa q (d_g+d)}}{\sinh{\kappa q(2d_g+d)}},\\
  c_2=-\frac{\sinh{\kappa q d_g}}{\sinh{\kappa q(2d_g+d)}}.
\end{gather*}
Thus the intra- and inter-layer interactions are 
\begin{subequations}
  \begin{gather}
    V_{intra}(\bm{q})=e\varphi_{\bm{q}}(d+d_g)=\frac{4\pi e^2}{\epsilon q}\frac{\sinh{\kappa q d_g}\sinh{\kappa q(d+d_g)}}{\sinh{\kappa q(2d_g+d)}},\\
    V_{inter}(\bm{q})=e\varphi_{\bm{q}}(d_g)=\frac{4\pi e^2}{\epsilon q}\frac{\sinh^2{\kappa q d_g}}{\sinh{\kappa q(2d_g+d)}}.
  \end{gather}
\end{subequations}
Expanded in exponentials, the interactions are approximated by 
\begin{subequations}
  \begin{gather}
    V_{intra}(\bm{q})\approx \frac{2\pi e^2}{\epsilon q}(1-\re^{-2\kappa q d_g}),\\
    V_{inter}(\bm{q})\approx \frac{2\pi e^2}{\epsilon q}(1-\re^{-2\kappa q d_g})\re^{-\kappa q d}.
  \end{gather}\label{eq:gate_screen_interaction}
\end{subequations}
From the expression in Eq. \eqref{eq:gate_screen_interaction} we can see that the screening mainly happens at the long-range part ($q\to 0$) of the interacting.

\section{Many-body Hamiltonian under Landau Level Basis}\label{app:many-body_landau_level}
In this section, we will derive the LL representations of the many-body Hamiltonian.
When magnetic field is applied, one should replace the kinetic momentum $\hbar\bm{k}$ in Eq. \eqref{eq:single-particle} by the canonical momentum $\bm{\Pi}=\hbar\bm{k}+e\bm{A}$ ($e=|e|$) according to Peierls substitution\cite{Brown1964}.
In Landau gauge $\bm{A}=(-yB,0)$, the wavefunction of LLs are 
\begin{equation}
  \phi_{nk_x}(\bm{r})=\frac{1}{\sqrt{L_x\ell}}\re^{ik_xx}\psi_{n}(y/\ell-\ell k_x),\; k_x\in[0, L_y/\ell^2],
\end{equation}
where $L_i$ is the system size in $i$ direction, $\ell=\sqrt{\hbar/eB}$ is the magnetic length and 
\begin{equation}
  \psi_{n}(x)=(2^nn!\sqrt{\pi})^{-1/2}\re^{-x^2/2}H_n(x)
\end{equation}
is the $n$-th level of 1D quantum Harmonic oscillator.
The LLs are complete and orthonormal, i.e. $\inp{\phi_{nk_x}}{\phi_{mk_x'}}=\delta_{nm}\delta_{k_xk_x'}$, $\sum_{nk_x}\ket{\phi_{nk_x}}\bra{\phi_{nk_x}}=\mathbbm{1}$.
Besides, it satisfies 
\begin{equation}
  \frac{\Pi^2}{2m_s}\ket{\phi_{nk_x}}=\hbar\omega_s(n+1/2)\ket{\phi_{nk_x}},
\end{equation}
where $\omega_s=eB/m_s$ is the cyclotron frequency.

It's easy to verify that 
\begin{subequations}
  \begin{gather*}
    \bra{e,\phi_{nk_x}}h^0_{\bm{\Pi}}\ket{e,\phi_{mk_x'}}=[\hbar\omega_{e}(n+1/2)-\mu_{ex}]\delta_{nm}\delta_{k_xk_x'},\\
    \bra{h,\phi_{nk_x}}h^0_{\bm{\Pi}}\ket{h,\phi_{mk_x'}}=-\hbar\omega_{h}(n+1/2)\delta_{nm}\delta_{k_xk_x'},\\
    \bra{e,\phi_{nk_x}}h^0_{\bm{\Pi}}\ket{h,\phi_{mk_x'}}=t\delta_{nm}\delta_{k_xk_x'}.
  \end{gather*}
\end{subequations}
Thus the single-particle part expressed under LL basis is written as 
\begin{align}
  H_0=&\sum_{ss'nk_x}\bra{s,\phi_{nk_x}}h^0_{\bm{\Pi}}-\mu\ket{s',\phi_{nk_x}}l^{\dagger}_{snk_x}l_{snk_x}\nonumber\\
  =&\sum_{ss'nk_x}(h^0_{n,ss'}-\mu\delta_{ss'})l^{\dagger}_{snk_x}l_{snk_x},\label{eq:single_particle_LL}
\end{align}
where $l^{\dagger}_{sn_x}$ is the creation operator for LL electrons and
\begin{equation}
  h^0_n=\begin{bmatrix}
    \hbar\omega_e(n+1/2)-\mu_{ex} & t \\
    t^* & -\hbar\omega_h(n+1/2)
  \end{bmatrix}.
\end{equation}

Use the relation 
\begin{equation}
  c_{s\bm{k}}=\sum_{nk_x'}\inp{\bm{k}}{\phi_{nk_x'}}l_{snk_x'},
\end{equation}
the interaction part Eq. \eqref{eq:interaction_plane_wave} becomes
\begin{align}
  H_{I}=&\frac{1}{2\mathcal{V}}\sum_{ss'}\sum_{\bm{k}_1'\bm{k}_2'\bm{q}}\sum_{n_ik_i}V_{ss'}(\bm{q})l^{\dagger}_{sn_1k_1}l^{\dagger}_{s'n_2k_2}l_{s'n_3k_3}l_{sn_4k_4}\nonumber\\
  \times&\inp{\phi_{n_1k_1}}{\bm{k}_1'}\inp{\bm{k}_1'-\bm{q}}{\phi_{n_4k_4}}\inp{\phi_{n_2k_2}}{\bm{k}_2'}\inp{\bm{k}_2'+\bm{q}}{\phi_{n_3k_3}}\nonumber\\
  =&\frac{1}{2\mathcal{V}}\sum_{ss'n_ik_i}\sum_{\bm{q}}V_{ss'}(\bm{q})l^{\dagger}_{sn_1k_1}l^{\dagger}_{s'n_2k_2}l_{s'n_3k_3}l_{sn_4k_4}\nonumber\\
  \times&\bra{\phi_{n_1k_1}}\re^{i\bm{q}\cdot\bm{r}}\ket{\phi_{n_4k_4}}\bra{\phi_{n_2k_2}}\re^{-i\bm{q}\cdot\bm{r}}\ket{\phi_{n_3k_3}}.\label{eq:manybody_LL_0}
\end{align}
To get the last equal, we use the identity
\begin{align}
  \sum_{\bm{k}}\ket{\bm{k}}\bra{\bm{k}-\bm{q}}=&\sum_{\bm{k}}\int\rd\bm{r}\ket{\bm{k}}\inp{\bm{k}-\bm{q}}{\bm{r}}\bra{\bm{r}}\nonumber\\
  =&\sum_{\bm{k}}\int\rd\bm{r}\ket{\bm{k}}\re^{-i(\bm{k}-\bm{q})\cdot\bm{r}}\bra{\bm{r}}\nonumber\\
  =&\sum_{\bm{k}}\int\rd\bm{r}\ket{\bm{k}}\bra{\bm{k}}\re^{i\bm{q}\cdot\bm{r}}\ket{\bm{r}}\bra{\bm{r}}\nonumber\\
  =&\re^{i\bm{q}\cdot\bm{r}}.\label{eq:identity}
\end{align}
Notice that $\bra{\phi_{n_2k_2}}\re^{-i\bm{q}\cdot\bm{r}}\ket{\phi_{n_3k_3}}\propto \int\rd x\;\re^{-i(k_2+q_x-k_3)x}\propto \delta_{k_3-k_2,q_x}$, Eq. \eqref{eq:manybody_LL_0} is finally simplified to
\begin{align}
  H_I=&\frac{1}{2\mathcal{V}}\sum_{ss'n_ik_i}\sum_{\bm{q}}V_{ss'}(\bm{q})\re^{iq_y(k_1-k_2)\ell^2}\Lambda_{n_4n_1}^*(\bm{q})\Lambda_{n_2n_3}(\bm{q})\nonumber\\
  &\times l^{\dagger}_{sn_1k_1+q_x/2}l^{\dagger}_{s'n_2k_2-q_x/2}l_{s'n_3k_2+q_x/2}l_{sn_4k_1-q_x/2},
\end{align}
where $\Lambda_{mn}(\bm{q})$ is the form factor for LLs
\begin{align}
  \Lambda_{mn}(\bm{q})\equiv&\bra{\phi_{mk-q_x/2}}\re^{-i\bm{q}\cdot{\bm{r}}}\ket{\phi_{nk+q_x/2}}\re^{ikq_y\ell^2}\nonumber\\
  =&\int\rd y\; \re^{-iq_y \ell y}\psi_{m}(y+q_x\ell/2)\psi_{n}(y-q_x\ell/2).
\end{align}
For $m\ge n$, $\Lambda_{mn}(\bm{q})$ is evaluated as\cite{macdonaldInfluenceLandaulevelMixing1984}
\begin{equation}
  \Lambda_{mn}(\bm{q})=\re^{-\frac{q^2\ell^2}{4}}\sqrt{\frac{m!}{n!}}\left(\frac{q_{-}\ell}{\sqrt{2}}\right)^{m-n}L_n^{(m-n)}\left(\frac{q^2\ell^2}{2}\right),\label{eq:form_factor}
\end{equation}
where $q_{-}=q_x-iq_y$ and $L_{n}^{(\alpha)}(x)$ is the Laguerre polynomial;
for $m<n$, $\Lambda_{mn}(\bm{q})$ could be got by $\Lambda_{mn}(\bm{q})=\Lambda^*_{nm}(-\bm{q})$.

\section{Mean-field Channels in Magnetic Field}\label{app:mean_field_channels}
As discussed in the main text, by requiring magnetic translation symmetry, the density matrix
\begin{equation}
  \rho_{sn_1,s'n_2}(k_x)\equiv\ave{l^{\dagger}_{s'n_2k_x}l_{sn_1k_x}}-\delta_{ss'}\delta_{sh}
\end{equation}
is $k_x$-independent.
Under Hartree Fock approximation, the mean-field Fock Hamiltonian is decoupled as 
\begin{align}
  H_F=&-\frac{1}{2\mathcal{V}}\sum_{ss'n_ik}\sum_{\bm{q}}V_{ss'}(\bm{q})\Lambda_{n_4n_1}^*(\bm{q})\Lambda_{n_2n_3}(\bm{q})\nonumber\\
  &\times\left[l^{\dagger}_{s'n_2k}l_{sn_4k}\rho_{s'n_3,sn_1}+l^{\dagger}_{sn_1k}l_{s'n_3k}\rho_{sn_4,s'n_2} \right].\label{eq:fock_hamilt_0}
\end{align}
According to Eq. \eqref{eq:form_factor}, we have $\Lambda_{mn}(\bm{q})\propto \re^{-i(m-n)\theta_{\bm{q}}}$.
Thus the $\bm{q}$ summation in Eq. \eqref{eq:fock_hamilt_0} is nonzero only when $n_4-n_1=n_2-n_3$.
It's convenient to define $M=n_3-n_1=n_2-n_4$ which labels independent condensation channels.
For condensation channel labeled by $M$, the only surviving density matrix elements are $\rho_{s'M+n,sn}$ and $\rho_{sn,s'n+M}$.

We argue that the index $M$ is just the angular momentum of exciton condensation.
In the absence of a magnetic field, the density matrix for the exciton condensation of angular momentum $M$ takes the form
\begin{equation}
  \rho_{eh\bm{k}}\sim k_{+}^{M}f(k^2),
\end{equation}
where $k_{+}=k_x+ik_y$ and $f$ is some analytic function.
After Peierls substitution and projecting to LL basis, we have 
\begin{align}
  &\bra{e,\phi_{nk_x}}\rho_{\bm{\Pi}}\ket{h,\phi_{mk_x'}}\nonumber\\
  \sim &\bra{\phi_{nk_x}}\Pi_{+}^{M}f(\Pi^2/\hbar^2)\ket{\phi_{mk_x'}}\nonumber\\
  \sim &\bra{\phi_{nk_x}}\Pi_{+}^{M}\ket{\phi_{mk_x'}}f((2m+1)eB/\hbar)\nonumber\\
  \sim& \delta_{n,m+M}\delta_{k_xk_x'}.
\end{align}
Due to Hermiticity, $\bra{h,\phi_{nk_x}}\rho_{\bm{\Pi}}\ket{e,\phi_{mk_x'}}\sim \delta_{n+M,m}\delta_{k_xk_x'}$.
Thus, for exciton condensation of angular momentum $M$, the only surviving EI order parameters under LL basis are $\rho_{en+M,hn}$ and $\rho_{hn,en+M}$.

\bibliography{ref.bib}

\begin{thebibliography}{41}%
\makeatletter
\providecommand \@ifxundefined [1]{%
 \@ifx{#1\undefined}
}%
\providecommand \@ifnum [1]{%
 \ifnum #1\expandafter \@firstoftwo
 \else \expandafter \@secondoftwo
 \fi
}%
\providecommand \@ifx [1]{%
 \ifx #1\expandafter \@firstoftwo
 \else \expandafter \@secondoftwo
 \fi
}%
\providecommand \natexlab [1]{#1}%
\providecommand \enquote  [1]{``#1''}%
\providecommand \bibnamefont  [1]{#1}%
\providecommand \bibfnamefont [1]{#1}%
\providecommand \citenamefont [1]{#1}%
\providecommand \href@noop [0]{\@secondoftwo}%
\providecommand \href [0]{\begingroup \@sanitize@url \@href}%
\providecommand \@href[1]{\@@startlink{#1}\@@href}%
\providecommand \@@href[1]{\endgroup#1\@@endlink}%
\providecommand \@sanitize@url [0]{\catcode `\\12\catcode `\$12\catcode
  `\&12\catcode `\#12\catcode `\^12\catcode `\_12\catcode `\%12\relax}%
\providecommand \@@startlink[1]{}%
\providecommand \@@endlink[0]{}%
\providecommand \url  [0]{\begingroup\@sanitize@url \@url }%
\providecommand \@url [1]{\endgroup\@href {#1}{\urlprefix }}%
\providecommand \urlprefix  [0]{URL }%
\providecommand \Eprint [0]{\href }%
\providecommand \doibase [0]{https://doi.org/}%
\providecommand \selectlanguage [0]{\@gobble}%
\providecommand \bibinfo  [0]{\@secondoftwo}%
\providecommand \bibfield  [0]{\@secondoftwo}%
\providecommand \translation [1]{[#1]}%
\providecommand \BibitemOpen [0]{}%
\providecommand \bibitemStop [0]{}%
\providecommand \bibitemNoStop [0]{.\EOS\space}%
\providecommand \EOS [0]{\spacefactor3000\relax}%
\providecommand \BibitemShut  [1]{\csname bibitem#1\endcsname}%
\let\auto@bib@innerbib\@empty
\bibitem [{\citenamefont {Butov}\ \emph {et~al.}(1994)\citenamefont {Butov},
  \citenamefont {Zrenner}, \citenamefont {Abstreiter}, \citenamefont
  {B{\"o}hm},\ and\ \citenamefont
  {Weimann}}]{butovCondensationIndirectExcitons1994}%
  \BibitemOpen
  \bibfield  {author} {\bibinfo {author} {\bibfnamefont {L.~V.}\ \bibnamefont
  {Butov}}, \bibinfo {author} {\bibfnamefont {A.}~\bibnamefont {Zrenner}},
  \bibinfo {author} {\bibfnamefont {G.}~\bibnamefont {Abstreiter}}, \bibinfo
  {author} {\bibfnamefont {G.}~\bibnamefont {B{\"o}hm}},\ and\ \bibinfo
  {author} {\bibfnamefont {G.}~\bibnamefont {Weimann}},\ }\href
  {https://doi.org/10.1103/PhysRevLett.73.304} {\bibfield  {journal} {\bibinfo
  {journal} {Physical Review Letters}\ }\textbf {\bibinfo {volume} {73}},\
  \bibinfo {pages} {304} (\bibinfo {year} {1994})}\BibitemShut {NoStop}%
\bibitem [{\citenamefont {Zhu}\ \emph {et~al.}(1995)\citenamefont {Zhu},
  \citenamefont {Littlewood}, \citenamefont {Hybertsen},\ and\ \citenamefont
  {Rice}}]{zhuExcitonCondensateSemiconductor1995}%
  \BibitemOpen
  \bibfield  {author} {\bibinfo {author} {\bibfnamefont {X.}~\bibnamefont
  {Zhu}}, \bibinfo {author} {\bibfnamefont {P.~B.}\ \bibnamefont {Littlewood}},
  \bibinfo {author} {\bibfnamefont {M.~S.}\ \bibnamefont {Hybertsen}},\ and\
  \bibinfo {author} {\bibfnamefont {T.~M.}\ \bibnamefont {Rice}},\ }\href
  {https://doi.org/10.1103/PhysRevLett.74.1633} {\bibfield  {journal} {\bibinfo
   {journal} {Physical Review Letters}\ }\textbf {\bibinfo {volume} {74}},\
  \bibinfo {pages} {1633} (\bibinfo {year} {1995})}\BibitemShut {NoStop}%
\bibitem [{\citenamefont {Littlewood}\ and\ \citenamefont
  {Zhu}(1996)}]{littlewoodPossibilitiesExcitonCondensation1996}%
  \BibitemOpen
  \bibfield  {author} {\bibinfo {author} {\bibfnamefont {P.~B.}\ \bibnamefont
  {Littlewood}}\ and\ \bibinfo {author} {\bibfnamefont {X.}~\bibnamefont
  {Zhu}},\ }\href {https://doi.org/10.1088/0031-8949/1996/T68/008} {\bibfield
  {journal} {\bibinfo  {journal} {Physica Scripta}\ }\textbf {\bibinfo {volume}
  {1996}},\ \bibinfo {pages} {56} (\bibinfo {year} {1996})}\BibitemShut
  {NoStop}%
\bibitem [{\citenamefont {Butov}(2003)}]{butovExcitonCondensationCoupled2003}%
  \BibitemOpen
  \bibfield  {author} {\bibinfo {author} {\bibfnamefont {L.~V.}\ \bibnamefont
  {Butov}},\ }\href {https://doi.org/10.1016/S0038-1098(03)00312-0} {\bibfield
  {journal} {\bibinfo  {journal} {Solid State Communications}\ }\bibinfo
  {series} {Quantum {{Phases}} at the {{Nanoscale}}},\ \textbf {\bibinfo
  {volume} {127}},\ \bibinfo {pages} {89} (\bibinfo {year} {2003})}\BibitemShut
  {NoStop}%
\bibitem [{\citenamefont {Littlewood}\ \emph {et~al.}(2004)\citenamefont
  {Littlewood}, \citenamefont {Eastham}, \citenamefont {Keeling}, \citenamefont
  {Marchetti}, \citenamefont {Simons},\ and\ \citenamefont
  {Szymanska}}]{littlewoodModelsCoherentExciton2004}%
  \BibitemOpen
  \bibfield  {author} {\bibinfo {author} {\bibfnamefont {P.~B.}\ \bibnamefont
  {Littlewood}}, \bibinfo {author} {\bibfnamefont {P.~R.}\ \bibnamefont
  {Eastham}}, \bibinfo {author} {\bibfnamefont {J.~M.~J.}\ \bibnamefont
  {Keeling}}, \bibinfo {author} {\bibfnamefont {F.~M.}\ \bibnamefont
  {Marchetti}}, \bibinfo {author} {\bibfnamefont {B.~D.}\ \bibnamefont
  {Simons}},\ and\ \bibinfo {author} {\bibfnamefont {M.~H.}\ \bibnamefont
  {Szymanska}},\ }\href {https://doi.org/10.1088/0953-8984/16/35/003}
  {\bibfield  {journal} {\bibinfo  {journal} {Journal of Physics: Condensed
  Matter}\ }\textbf {\bibinfo {volume} {16}},\ \bibinfo {pages} {S3597}
  (\bibinfo {year} {2004})}\BibitemShut {NoStop}%
\bibitem [{\citenamefont {Min}\ \emph {et~al.}(2008)\citenamefont {Min},
  \citenamefont {Bistritzer}, \citenamefont {Su},\ and\ \citenamefont
  {MacDonald}}]{minRoomtemperatureSuperfluidityGraphene2008}%
  \BibitemOpen
  \bibfield  {author} {\bibinfo {author} {\bibfnamefont {H.}~\bibnamefont
  {Min}}, \bibinfo {author} {\bibfnamefont {R.}~\bibnamefont {Bistritzer}},
  \bibinfo {author} {\bibfnamefont {J.-J.}\ \bibnamefont {Su}},\ and\ \bibinfo
  {author} {\bibfnamefont {A.~H.}\ \bibnamefont {MacDonald}},\ }\href
  {https://doi.org/10.1103/PhysRevB.78.121401} {\bibfield  {journal} {\bibinfo
  {journal} {Physical Review B}\ }\textbf {\bibinfo {volume} {78}},\ \bibinfo
  {pages} {121401} (\bibinfo {year} {2008})}\BibitemShut {NoStop}%
\bibitem [{\citenamefont {Fogler}\ \emph {et~al.}(2014)\citenamefont {Fogler},
  \citenamefont {Butov},\ and\ \citenamefont
  {Novoselov}}]{foglerHightemperatureSuperfluidityIndirect2014a}%
  \BibitemOpen
  \bibfield  {author} {\bibinfo {author} {\bibfnamefont {M.~M.}\ \bibnamefont
  {Fogler}}, \bibinfo {author} {\bibfnamefont {L.~V.}\ \bibnamefont {Butov}},\
  and\ \bibinfo {author} {\bibfnamefont {K.~S.}\ \bibnamefont {Novoselov}},\
  }\href {https://doi.org/10.1038/ncomms5555} {\bibfield  {journal} {\bibinfo
  {journal} {Nature Communications}\ }\textbf {\bibinfo {volume} {5}},\
  \bibinfo {pages} {4555} (\bibinfo {year} {2014})}\BibitemShut {NoStop}%
\bibitem [{\citenamefont {Wu}\ \emph {et~al.}(2015)\citenamefont {Wu},
  \citenamefont {Xue},\ and\ \citenamefont
  {MacDonald}}]{wuTheoryTwodimensionalSpatially2015}%
  \BibitemOpen
  \bibfield  {author} {\bibinfo {author} {\bibfnamefont {F.-C.}\ \bibnamefont
  {Wu}}, \bibinfo {author} {\bibfnamefont {F.}~\bibnamefont {Xue}},\ and\
  \bibinfo {author} {\bibfnamefont {A.~H.}\ \bibnamefont {MacDonald}},\ }\href
  {https://doi.org/10.1103/PhysRevB.92.165121} {\bibfield  {journal} {\bibinfo
  {journal} {Physical Review B}\ }\textbf {\bibinfo {volume} {92}},\ \bibinfo
  {pages} {165121} (\bibinfo {year} {2015})}\BibitemShut {NoStop}%
\bibitem [{\citenamefont {High}\ \emph {et~al.}(2012)\citenamefont {High},
  \citenamefont {Leonard}, \citenamefont {Hammack}, \citenamefont {Fogler},
  \citenamefont {Butov}, \citenamefont {Kavokin}, \citenamefont {Campman},\
  and\ \citenamefont {Gossard}}]{highSpontaneousCoherenceCold2012}%
  \BibitemOpen
  \bibfield  {author} {\bibinfo {author} {\bibfnamefont {A.~A.}\ \bibnamefont
  {High}}, \bibinfo {author} {\bibfnamefont {J.~R.}\ \bibnamefont {Leonard}},
  \bibinfo {author} {\bibfnamefont {A.~T.}\ \bibnamefont {Hammack}}, \bibinfo
  {author} {\bibfnamefont {M.~M.}\ \bibnamefont {Fogler}}, \bibinfo {author}
  {\bibfnamefont {L.~V.}\ \bibnamefont {Butov}}, \bibinfo {author}
  {\bibfnamefont {A.~V.}\ \bibnamefont {Kavokin}}, \bibinfo {author}
  {\bibfnamefont {K.~L.}\ \bibnamefont {Campman}},\ and\ \bibinfo {author}
  {\bibfnamefont {A.~C.}\ \bibnamefont {Gossard}},\ }\href
  {https://doi.org/10.1038/nature10903} {\bibfield  {journal} {\bibinfo
  {journal} {Nature}\ }\textbf {\bibinfo {volume} {483}},\ \bibinfo {pages}
  {584} (\bibinfo {year} {2012})}\BibitemShut {NoStop}%
\bibitem [{\citenamefont {Li}\ \emph {et~al.}(2017)\citenamefont {Li},
  \citenamefont {Taniguchi}, \citenamefont {Watanabe}, \citenamefont {Hone},\
  and\ \citenamefont {Dean}}]{liExcitonicSuperfluidPhase2017}%
  \BibitemOpen
  \bibfield  {author} {\bibinfo {author} {\bibfnamefont {J.~I.~A.}\
  \bibnamefont {Li}}, \bibinfo {author} {\bibfnamefont {T.}~\bibnamefont
  {Taniguchi}}, \bibinfo {author} {\bibfnamefont {K.}~\bibnamefont {Watanabe}},
  \bibinfo {author} {\bibfnamefont {J.}~\bibnamefont {Hone}},\ and\ \bibinfo
  {author} {\bibfnamefont {C.~R.}\ \bibnamefont {Dean}},\ }\href
  {https://doi.org/10.1038/nphys4140} {\bibfield  {journal} {\bibinfo
  {journal} {Nature Physics}\ }\textbf {\bibinfo {volume} {13}},\ \bibinfo
  {pages} {751} (\bibinfo {year} {2017})}\BibitemShut {NoStop}%
\bibitem [{\citenamefont {Du}\ \emph {et~al.}(2017)\citenamefont {Du},
  \citenamefont {Li}, \citenamefont {Lou}, \citenamefont {Sullivan},
  \citenamefont {Chang}, \citenamefont {Kono},\ and\ \citenamefont
  {Du}}]{duEvidenceTopologicalExcitonic2017}%
  \BibitemOpen
  \bibfield  {author} {\bibinfo {author} {\bibfnamefont {L.}~\bibnamefont
  {Du}}, \bibinfo {author} {\bibfnamefont {X.}~\bibnamefont {Li}}, \bibinfo
  {author} {\bibfnamefont {W.}~\bibnamefont {Lou}}, \bibinfo {author}
  {\bibfnamefont {G.}~\bibnamefont {Sullivan}}, \bibinfo {author}
  {\bibfnamefont {K.}~\bibnamefont {Chang}}, \bibinfo {author} {\bibfnamefont
  {J.}~\bibnamefont {Kono}},\ and\ \bibinfo {author} {\bibfnamefont {R.-R.}\
  \bibnamefont {Du}},\ }\href {https://doi.org/10.1038/s41467-017-01988-1}
  {\bibfield  {journal} {\bibinfo  {journal} {Nature Communications}\ }\textbf
  {\bibinfo {volume} {8}},\ \bibinfo {pages} {1971} (\bibinfo {year}
  {2017})}\BibitemShut {NoStop}%
\bibitem [{\citenamefont {Wang}\ \emph {et~al.}(2018)\citenamefont {Wang},
  \citenamefont {Chiu}, \citenamefont {Honz}, \citenamefont {Mak},\ and\
  \citenamefont {Shan}}]{wangElectricalTuningInterlayer2018}%
  \BibitemOpen
  \bibfield  {author} {\bibinfo {author} {\bibfnamefont {Z.}~\bibnamefont
  {Wang}}, \bibinfo {author} {\bibfnamefont {Y.-H.}\ \bibnamefont {Chiu}},
  \bibinfo {author} {\bibfnamefont {K.}~\bibnamefont {Honz}}, \bibinfo {author}
  {\bibfnamefont {K.~F.}\ \bibnamefont {Mak}},\ and\ \bibinfo {author}
  {\bibfnamefont {J.}~\bibnamefont {Shan}},\ }\href
  {https://doi.org/10.1021/acs.nanolett.7b03667} {\bibfield  {journal}
  {\bibinfo  {journal} {Nano Letters}\ }\textbf {\bibinfo {volume} {18}},\
  \bibinfo {pages} {137} (\bibinfo {year} {2018})}\BibitemShut {NoStop}%
\bibitem [{\citenamefont {Xie}\ and\ \citenamefont
  {MacDonald}(2018)}]{xieElectricalReservoirsBilayer2018}%
  \BibitemOpen
  \bibfield  {author} {\bibinfo {author} {\bibfnamefont {M.}~\bibnamefont
  {Xie}}\ and\ \bibinfo {author} {\bibfnamefont {A.~H.}\ \bibnamefont
  {MacDonald}},\ }\href {https://doi.org/10.1103/PhysRevLett.121.067702}
  {\bibfield  {journal} {\bibinfo  {journal} {Physical Review Letters}\
  }\textbf {\bibinfo {volume} {121}},\ \bibinfo {pages} {067702} (\bibinfo
  {year} {2018})}\BibitemShut {NoStop}%
\bibitem [{\citenamefont {Wang}\ \emph {et~al.}(2019)\citenamefont {Wang},
  \citenamefont {Rhodes}, \citenamefont {Watanabe}, \citenamefont {Taniguchi},
  \citenamefont {Hone}, \citenamefont {Shan},\ and\ \citenamefont
  {Mak}}]{wangEvidenceHightemperatureExciton2019a}%
  \BibitemOpen
  \bibfield  {author} {\bibinfo {author} {\bibfnamefont {Z.}~\bibnamefont
  {Wang}}, \bibinfo {author} {\bibfnamefont {D.~A.}\ \bibnamefont {Rhodes}},
  \bibinfo {author} {\bibfnamefont {K.}~\bibnamefont {Watanabe}}, \bibinfo
  {author} {\bibfnamefont {T.}~\bibnamefont {Taniguchi}}, \bibinfo {author}
  {\bibfnamefont {J.~C.}\ \bibnamefont {Hone}}, \bibinfo {author}
  {\bibfnamefont {J.}~\bibnamefont {Shan}},\ and\ \bibinfo {author}
  {\bibfnamefont {K.~F.}\ \bibnamefont {Mak}},\ }\href
  {https://doi.org/10.1038/s41586-019-1591-7} {\bibfield  {journal} {\bibinfo
  {journal} {Nature}\ }\textbf {\bibinfo {volume} {574}},\ \bibinfo {pages}
  {76} (\bibinfo {year} {2019})}\BibitemShut {NoStop}%
\bibitem [{\citenamefont {Ma}\ \emph {et~al.}(2021)\citenamefont {Ma},
  \citenamefont {Nguyen}, \citenamefont {Wang}, \citenamefont {Zeng},
  \citenamefont {Watanabe}, \citenamefont {Taniguchi}, \citenamefont
  {MacDonald}, \citenamefont {Mak},\ and\ \citenamefont
  {Shan}}]{maStronglyCorrelatedExcitonic2021}%
  \BibitemOpen
  \bibfield  {author} {\bibinfo {author} {\bibfnamefont {L.}~\bibnamefont
  {Ma}}, \bibinfo {author} {\bibfnamefont {P.~X.}\ \bibnamefont {Nguyen}},
  \bibinfo {author} {\bibfnamefont {Z.}~\bibnamefont {Wang}}, \bibinfo {author}
  {\bibfnamefont {Y.}~\bibnamefont {Zeng}}, \bibinfo {author} {\bibfnamefont
  {K.}~\bibnamefont {Watanabe}}, \bibinfo {author} {\bibfnamefont
  {T.}~\bibnamefont {Taniguchi}}, \bibinfo {author} {\bibfnamefont {A.~H.}\
  \bibnamefont {MacDonald}}, \bibinfo {author} {\bibfnamefont {K.~F.}\
  \bibnamefont {Mak}},\ and\ \bibinfo {author} {\bibfnamefont {J.}~\bibnamefont
  {Shan}},\ }\href {https://doi.org/10.1038/s41586-021-03947-9} {\bibfield
  {journal} {\bibinfo  {journal} {Nature}\ }\textbf {\bibinfo {volume} {598}},\
  \bibinfo {pages} {585} (\bibinfo {year} {2021})}\BibitemShut {NoStop}%
\bibitem [{\citenamefont {J{\'e}rome}\ \emph {et~al.}(1967)\citenamefont
  {J{\'e}rome}, \citenamefont {Rice},\ and\ \citenamefont {Kohn}}]{Jerome1967}%
  \BibitemOpen
  \bibfield  {author} {\bibinfo {author} {\bibfnamefont {D.}~\bibnamefont
  {J{\'e}rome}}, \bibinfo {author} {\bibfnamefont {T.~M.}\ \bibnamefont
  {Rice}},\ and\ \bibinfo {author} {\bibfnamefont {W.}~\bibnamefont {Kohn}},\
  }\href {https://doi.org/10.1103/PhysRev.158.462} {\bibfield  {journal}
  {\bibinfo  {journal} {Physical Review}\ }\textbf {\bibinfo {volume} {158}},\
  \bibinfo {pages} {462} (\bibinfo {year} {1967})}\BibitemShut {NoStop}%
\bibitem [{\citenamefont {Eisenstein}\ and\ \citenamefont
  {MacDonald}(2004)}]{eisensteinBoseEinsteinCondensation2004}%
  \BibitemOpen
  \bibfield  {author} {\bibinfo {author} {\bibfnamefont {J.~P.}\ \bibnamefont
  {Eisenstein}}\ and\ \bibinfo {author} {\bibfnamefont {A.~H.}\ \bibnamefont
  {MacDonald}},\ }\href {https://doi.org/10.1038/nature03081} {\bibfield
  {journal} {\bibinfo  {journal} {Nature}\ }\textbf {\bibinfo {volume} {432}},\
  \bibinfo {pages} {691} (\bibinfo {year} {2004})}\BibitemShut {NoStop}%
\bibitem [{\citenamefont {Nandi}\ \emph {et~al.}(2012)\citenamefont {Nandi},
  \citenamefont {Finck}, \citenamefont {Eisenstein}, \citenamefont {Pfeiffer},\
  and\ \citenamefont {West}}]{nandiExcitonCondensationPerfect2012}%
  \BibitemOpen
  \bibfield  {author} {\bibinfo {author} {\bibfnamefont {D.}~\bibnamefont
  {Nandi}}, \bibinfo {author} {\bibfnamefont {A.~D.~K.}\ \bibnamefont {Finck}},
  \bibinfo {author} {\bibfnamefont {J.~P.}\ \bibnamefont {Eisenstein}},
  \bibinfo {author} {\bibfnamefont {L.~N.}\ \bibnamefont {Pfeiffer}},\ and\
  \bibinfo {author} {\bibfnamefont {K.~W.}\ \bibnamefont {West}},\ }\href
  {https://doi.org/10.1038/nature11302} {\bibfield  {journal} {\bibinfo
  {journal} {Nature}\ }\textbf {\bibinfo {volume} {488}},\ \bibinfo {pages}
  {481} (\bibinfo {year} {2012})}\BibitemShut {NoStop}%
\bibitem [{\citenamefont {Zhu}\ \emph {et~al.}(2019)\citenamefont {Zhu},
  \citenamefont {Tu}, \citenamefont {Tong},\ and\ \citenamefont
  {Yao}}]{zhuGateTuningExciton2019}%
  \BibitemOpen
  \bibfield  {author} {\bibinfo {author} {\bibfnamefont {Q.}~\bibnamefont
  {Zhu}}, \bibinfo {author} {\bibfnamefont {M.~W.-Y.}\ \bibnamefont {Tu}},
  \bibinfo {author} {\bibfnamefont {Q.}~\bibnamefont {Tong}},\ and\ \bibinfo
  {author} {\bibfnamefont {W.}~\bibnamefont {Yao}},\ }\href
  {https://doi.org/10.1126/sciadv.aau6120} {\bibfield  {journal} {\bibinfo
  {journal} {Science Advances}\ }\textbf {\bibinfo {volume} {5}},\ \bibinfo
  {pages} {eaau6120} (\bibinfo {year} {2019})}\BibitemShut {NoStop}%
\bibitem [{\citenamefont {Tong}\ \emph {et~al.}(2017)\citenamefont {Tong},
  \citenamefont {Yu}, \citenamefont {Zhu}, \citenamefont {Wang}, \citenamefont
  {Xu},\ and\ \citenamefont {Yao}}]{tongTopologicalMosaicsMoire2017}%
  \BibitemOpen
  \bibfield  {author} {\bibinfo {author} {\bibfnamefont {Q.}~\bibnamefont
  {Tong}}, \bibinfo {author} {\bibfnamefont {H.}~\bibnamefont {Yu}}, \bibinfo
  {author} {\bibfnamefont {Q.}~\bibnamefont {Zhu}}, \bibinfo {author}
  {\bibfnamefont {Y.}~\bibnamefont {Wang}}, \bibinfo {author} {\bibfnamefont
  {X.}~\bibnamefont {Xu}},\ and\ \bibinfo {author} {\bibfnamefont
  {W.}~\bibnamefont {Yao}},\ }\href {https://doi.org/10.1038/nphys3968}
  {\bibfield  {journal} {\bibinfo  {journal} {Nature Physics}\ }\textbf
  {\bibinfo {volume} {13}},\ \bibinfo {pages} {356} (\bibinfo {year}
  {2017})}\BibitemShut {NoStop}%
\bibitem [{\citenamefont {Zee}(1995)}]{zeeQuantumHallFluids1995}%
  \BibitemOpen
  \bibfield  {author} {\bibinfo {author} {\bibfnamefont {A.}~\bibnamefont
  {Zee}}\ }(\bibinfo {year} {1995})\ pp.\ \bibinfo {pages} {99--153},\ \Eprint
  {https://arxiv.org/abs/cond-mat/9501022} {arxiv:cond-mat/9501022}
  \BibitemShut {NoStop}%
\bibitem [{\citenamefont {Zeng}\ and\ \citenamefont
  {MacDonald}(2020)}]{zengElectricallyControlledTwodimensional2020a}%
  \BibitemOpen
  \bibfield  {author} {\bibinfo {author} {\bibfnamefont {Y.}~\bibnamefont
  {Zeng}}\ and\ \bibinfo {author} {\bibfnamefont {A.~H.}\ \bibnamefont
  {MacDonald}},\ }\href {https://doi.org/10.1103/PhysRevB.102.085154}
  {\bibfield  {journal} {\bibinfo  {journal} {Physical Review B}\ }\textbf
  {\bibinfo {volume} {102}},\ \bibinfo {pages} {085154} (\bibinfo {year}
  {2020})}\BibitemShut {NoStop}%
\bibitem [{\citenamefont {Hallam}\ \emph {et~al.}(1996)\citenamefont {Hallam},
  \citenamefont {Weis},\ and\ \citenamefont
  {Maksym}}]{hallamScreeningElectronelectronInteraction1996}%
  \BibitemOpen
  \bibfield  {author} {\bibinfo {author} {\bibfnamefont {L.~D.}\ \bibnamefont
  {Hallam}}, \bibinfo {author} {\bibfnamefont {J.}~\bibnamefont {Weis}},\ and\
  \bibinfo {author} {\bibfnamefont {P.~A.}\ \bibnamefont {Maksym}},\ }\href
  {https://doi.org/10.1103/PhysRevB.53.1452} {\bibfield  {journal} {\bibinfo
  {journal} {Physical Review B}\ }\textbf {\bibinfo {volume} {53}},\ \bibinfo
  {pages} {1452} (\bibinfo {year} {1996})}\BibitemShut {NoStop}%
\bibitem [{\citenamefont {Shim}\ and\ \citenamefont
  {MacDonald}(2009)}]{shimSpinorbitInteractionsBilayer2009}%
  \BibitemOpen
  \bibfield  {author} {\bibinfo {author} {\bibfnamefont {Y.-P.}\ \bibnamefont
  {Shim}}\ and\ \bibinfo {author} {\bibfnamefont {A.~H.}\ \bibnamefont
  {MacDonald}},\ }\href {https://doi.org/10.1103/PhysRevB.79.235329} {\bibfield
   {journal} {\bibinfo  {journal} {Physical Review B}\ }\textbf {\bibinfo
  {volume} {79}},\ \bibinfo {pages} {235329} (\bibinfo {year}
  {2009})}\BibitemShut {NoStop}%
\bibitem [{\citenamefont
  {MacDonald}(1984)}]{macdonaldInfluenceLandaulevelMixing1984}%
  \BibitemOpen
  \bibfield  {author} {\bibinfo {author} {\bibfnamefont {A.~H.}\ \bibnamefont
  {MacDonald}},\ }\href {https://doi.org/10.1103/PhysRevB.30.4392} {\bibfield
  {journal} {\bibinfo  {journal} {Physical Review B}\ }\textbf {\bibinfo
  {volume} {30}},\ \bibinfo {pages} {4392} (\bibinfo {year}
  {1984})}\BibitemShut {NoStop}%
\bibitem [{\citenamefont {Zou}\ \emph {et~al.}(2023)\citenamefont {Zou},
  \citenamefont {Zeng}, \citenamefont {MacDonald},\ and\ \citenamefont
  {Strashko}}]{zouElectricalControlTwoDimensional2023}%
  \BibitemOpen
  \bibfield  {author} {\bibinfo {author} {\bibfnamefont {B.}~\bibnamefont
  {Zou}}, \bibinfo {author} {\bibfnamefont {Y.}~\bibnamefont {Zeng}}, \bibinfo
  {author} {\bibfnamefont {A.~H.}\ \bibnamefont {MacDonald}},\ and\ \bibinfo
  {author} {\bibfnamefont {A.}~\bibnamefont {Strashko}},\ }\href
  {https://doi.org/10.48550/arXiv.2309.04600} {\bibinfo {title} {Electrical
  {{Control}} of {{Two-Dimensional Electron-Hole Fluids}} in the {{Quantum Hall
  Regime}}}} (\bibinfo {year} {2023}),\ \Eprint
  {https://arxiv.org/abs/2309.04600} {arxiv:2309.04600 [cond-mat]} \BibitemShut
  {NoStop}%
\bibitem [{\citenamefont {Korm{\'a}nyos}\ \emph {et~al.}(2015)\citenamefont
  {Korm{\'a}nyos}, \citenamefont {Burkard}, \citenamefont {Gmitra},
  \citenamefont {Fabian}, \citenamefont {Z{\'o}lyomi}, \citenamefont
  {Drummond},\ and\ \citenamefont
  {Fal'ko}}]{kormanyosTheoryTwodimensionalTransition2015a}%
  \BibitemOpen
  \bibfield  {author} {\bibinfo {author} {\bibfnamefont {A.}~\bibnamefont
  {Korm{\'a}nyos}}, \bibinfo {author} {\bibfnamefont {G.}~\bibnamefont
  {Burkard}}, \bibinfo {author} {\bibfnamefont {M.}~\bibnamefont {Gmitra}},
  \bibinfo {author} {\bibfnamefont {J.}~\bibnamefont {Fabian}}, \bibinfo
  {author} {\bibfnamefont {V.}~\bibnamefont {Z{\'o}lyomi}}, \bibinfo {author}
  {\bibfnamefont {N.~D.}\ \bibnamefont {Drummond}},\ and\ \bibinfo {author}
  {\bibfnamefont {V.}~\bibnamefont {Fal'ko}},\ }\href
  {https://doi.org/10.1088/2053-1583/2/2/022001} {\bibfield  {journal}
  {\bibinfo  {journal} {2D Materials}\ }\textbf {\bibinfo {volume} {2}},\
  \bibinfo {pages} {022001} (\bibinfo {year} {2015})}\BibitemShut {NoStop}%
\bibitem [{\citenamefont {Cai}\ \emph {et~al.}(2007)\citenamefont {Cai},
  \citenamefont {Zhang}, \citenamefont {Zeng}, \citenamefont {Cheng},\ and\
  \citenamefont {Xu}}]{caiInfraredReflectanceSpectrum2007}%
  \BibitemOpen
  \bibfield  {author} {\bibinfo {author} {\bibfnamefont {Y.}~\bibnamefont
  {Cai}}, \bibinfo {author} {\bibfnamefont {L.}~\bibnamefont {Zhang}}, \bibinfo
  {author} {\bibfnamefont {Q.}~\bibnamefont {Zeng}}, \bibinfo {author}
  {\bibfnamefont {L.}~\bibnamefont {Cheng}},\ and\ \bibinfo {author}
  {\bibfnamefont {Y.}~\bibnamefont {Xu}},\ }\href
  {https://doi.org/10.1016/j.ssc.2006.10.040} {\bibfield  {journal} {\bibinfo
  {journal} {Solid State Communications}\ }\textbf {\bibinfo {volume} {141}},\
  \bibinfo {pages} {262} (\bibinfo {year} {2007})}\BibitemShut {NoStop}%
\bibitem [{\citenamefont {Filinov}\ \emph {et~al.}(2010)\citenamefont
  {Filinov}, \citenamefont {Prokof'ev},\ and\ \citenamefont
  {Bonitz}}]{filinovBerezinskiiKosterlitzThoulessTransitionTwoDimensional2010}%
  \BibitemOpen
  \bibfield  {author} {\bibinfo {author} {\bibfnamefont {A.}~\bibnamefont
  {Filinov}}, \bibinfo {author} {\bibfnamefont {N.~V.}\ \bibnamefont
  {Prokof'ev}},\ and\ \bibinfo {author} {\bibfnamefont {M.}~\bibnamefont
  {Bonitz}},\ }\href {https://doi.org/10.1103/PhysRevLett.105.070401}
  {\bibfield  {journal} {\bibinfo  {journal} {Physical Review Letters}\
  }\textbf {\bibinfo {volume} {105}},\ \bibinfo {pages} {070401} (\bibinfo
  {year} {2010})}\BibitemShut {NoStop}%
\bibitem [{\citenamefont
  {Shoenberg}(2009)}]{shoenbergMagneticOscillationsMetals2009}%
  \BibitemOpen
  \bibfield  {author} {\bibinfo {author} {\bibfnamefont {D.}~\bibnamefont
  {Shoenberg}},\ }\href@noop {} {\emph {\bibinfo {title} {Magnetic
  {{Oscillations}} in {{Metals}}}}}\ (\bibinfo  {publisher} {{Cambridge
  University Press}},\ \bibinfo {year} {2009})\BibitemShut {NoStop}%
\bibitem [{\citenamefont {Knolle}\ and\ \citenamefont
  {Cooper}(2015)}]{knolleQuantumOscillationsFermi2015}%
  \BibitemOpen
  \bibfield  {author} {\bibinfo {author} {\bibfnamefont {J.}~\bibnamefont
  {Knolle}}\ and\ \bibinfo {author} {\bibfnamefont {N.~R.}\ \bibnamefont
  {Cooper}},\ }\href {https://doi.org/10.1103/PhysRevLett.115.146401}
  {\bibfield  {journal} {\bibinfo  {journal} {Physical Review Letters}\
  }\textbf {\bibinfo {volume} {115}},\ \bibinfo {pages} {146401} (\bibinfo
  {year} {2015})}\BibitemShut {NoStop}%
\bibitem [{\citenamefont {Zhang}\ \emph {et~al.}(2016)\citenamefont {Zhang},
  \citenamefont {Song},\ and\ \citenamefont
  {Wang}}]{zhangQuantumOscillationNarrowGap2016}%
  \BibitemOpen
  \bibfield  {author} {\bibinfo {author} {\bibfnamefont {L.}~\bibnamefont
  {Zhang}}, \bibinfo {author} {\bibfnamefont {X.-Y.}\ \bibnamefont {Song}},\
  and\ \bibinfo {author} {\bibfnamefont {F.}~\bibnamefont {Wang}},\ }\href
  {https://doi.org/10.1103/PhysRevLett.116.046404} {\bibfield  {journal}
  {\bibinfo  {journal} {Physical Review Letters}\ }\textbf {\bibinfo {volume}
  {116}},\ \bibinfo {pages} {046404} (\bibinfo {year} {2016})}\BibitemShut
  {NoStop}%
\bibitem [{\citenamefont {Ong}(2018)}]{ongQuantumOscillationsInsulator2018}%
  \BibitemOpen
  \bibfield  {author} {\bibinfo {author} {\bibfnamefont {N.~P.}\ \bibnamefont
  {Ong}},\ }\href {https://doi.org/10.1126/science.aau3840} {\bibfield
  {journal} {\bibinfo  {journal} {Science}\ }\textbf {\bibinfo {volume}
  {362}},\ \bibinfo {pages} {32} (\bibinfo {year} {2018})}\BibitemShut
  {NoStop}%
\bibitem [{\citenamefont {Xiang}\ \emph {et~al.}(2018)\citenamefont {Xiang},
  \citenamefont {Kasahara}, \citenamefont {Asaba}, \citenamefont {Lawson},
  \citenamefont {Tinsman}, \citenamefont {Chen}, \citenamefont {Sugimoto},
  \citenamefont {Kawaguchi}, \citenamefont {Sato}, \citenamefont {Li},
  \citenamefont {Yao}, \citenamefont {Chen}, \citenamefont {Iga}, \citenamefont
  {Singleton}, \citenamefont {Matsuda},\ and\ \citenamefont
  {Li}}]{xiangQuantumOscillationsElectrical2018}%
  \BibitemOpen
  \bibfield  {author} {\bibinfo {author} {\bibfnamefont {Z.}~\bibnamefont
  {Xiang}}, \bibinfo {author} {\bibfnamefont {Y.}~\bibnamefont {Kasahara}},
  \bibinfo {author} {\bibfnamefont {T.}~\bibnamefont {Asaba}}, \bibinfo
  {author} {\bibfnamefont {B.}~\bibnamefont {Lawson}}, \bibinfo {author}
  {\bibfnamefont {C.}~\bibnamefont {Tinsman}}, \bibinfo {author} {\bibfnamefont
  {L.}~\bibnamefont {Chen}}, \bibinfo {author} {\bibfnamefont {K.}~\bibnamefont
  {Sugimoto}}, \bibinfo {author} {\bibfnamefont {S.}~\bibnamefont {Kawaguchi}},
  \bibinfo {author} {\bibfnamefont {Y.}~\bibnamefont {Sato}}, \bibinfo {author}
  {\bibfnamefont {G.}~\bibnamefont {Li}}, \bibinfo {author} {\bibfnamefont
  {S.}~\bibnamefont {Yao}}, \bibinfo {author} {\bibfnamefont {Y.~L.}\
  \bibnamefont {Chen}}, \bibinfo {author} {\bibfnamefont {F.}~\bibnamefont
  {Iga}}, \bibinfo {author} {\bibfnamefont {J.}~\bibnamefont {Singleton}},
  \bibinfo {author} {\bibfnamefont {Y.}~\bibnamefont {Matsuda}},\ and\ \bibinfo
  {author} {\bibfnamefont {L.}~\bibnamefont {Li}},\ }\href
  {https://doi.org/10.1126/science.aap9607} {\bibfield  {journal} {\bibinfo
  {journal} {Science}\ }\textbf {\bibinfo {volume} {362}},\ \bibinfo {pages}
  {65} (\bibinfo {year} {2018})}\BibitemShut {NoStop}%
\bibitem [{\citenamefont {Panda}\ \emph {et~al.}(2022)\citenamefont {Panda},
  \citenamefont {Banerjee},\ and\ \citenamefont
  {Randeria}}]{pandaQuantumOscillationsMagnetization2022}%
  \BibitemOpen
  \bibfield  {author} {\bibinfo {author} {\bibfnamefont {A.}~\bibnamefont
  {Panda}}, \bibinfo {author} {\bibfnamefont {S.}~\bibnamefont {Banerjee}},\
  and\ \bibinfo {author} {\bibfnamefont {M.}~\bibnamefont {Randeria}},\ }\href
  {https://doi.org/10.1073/pnas.2208373119} {\bibfield  {journal} {\bibinfo
  {journal} {Proceedings of the National Academy of Sciences}\ }\textbf
  {\bibinfo {volume} {119}},\ \bibinfo {pages} {e2208373119} (\bibinfo {year}
  {2022})}\BibitemShut {NoStop}%
\bibitem [{\citenamefont {Suzuki}\ \emph {et~al.}(2004)\citenamefont {Suzuki},
  \citenamefont {Takashina}, \citenamefont {Miyashita},\ and\ \citenamefont
  {Hirayama}}]{suzukiLandauLevelHybridizationQuantum2004}%
  \BibitemOpen
  \bibfield  {author} {\bibinfo {author} {\bibfnamefont {K.}~\bibnamefont
  {Suzuki}}, \bibinfo {author} {\bibfnamefont {K.}~\bibnamefont {Takashina}},
  \bibinfo {author} {\bibfnamefont {S.}~\bibnamefont {Miyashita}},\ and\
  \bibinfo {author} {\bibfnamefont {Y.}~\bibnamefont {Hirayama}},\ }\href
  {https://doi.org/10.1103/PhysRevLett.93.016803} {\bibfield  {journal}
  {\bibinfo  {journal} {Physical Review Letters}\ }\textbf {\bibinfo {volume}
  {93}},\ \bibinfo {pages} {016803} (\bibinfo {year} {2004})}\BibitemShut
  {NoStop}%
\bibitem [{\citenamefont {Tan}\ \emph {et~al.}(2015)\citenamefont {Tan},
  \citenamefont {Hsu}, \citenamefont {Zeng}, \citenamefont {Hatnean},
  \citenamefont {Harrison}, \citenamefont {Zhu}, \citenamefont {Hartstein},
  \citenamefont {Kiourlappou}, \citenamefont {Srivastava}, \citenamefont
  {Johannes}, \citenamefont {Murphy}, \citenamefont {Park}, \citenamefont
  {Balicas}, \citenamefont {Lonzarich}, \citenamefont {Balakrishnan},\ and\
  \citenamefont {Sebastian}}]{tanUnconventionalFermiSurface2015}%
  \BibitemOpen
  \bibfield  {author} {\bibinfo {author} {\bibfnamefont {B.~S.}\ \bibnamefont
  {Tan}}, \bibinfo {author} {\bibfnamefont {Y.-T.}\ \bibnamefont {Hsu}},
  \bibinfo {author} {\bibfnamefont {B.}~\bibnamefont {Zeng}}, \bibinfo {author}
  {\bibfnamefont {M.~C.}\ \bibnamefont {Hatnean}}, \bibinfo {author}
  {\bibfnamefont {N.}~\bibnamefont {Harrison}}, \bibinfo {author}
  {\bibfnamefont {Z.}~\bibnamefont {Zhu}}, \bibinfo {author} {\bibfnamefont
  {M.}~\bibnamefont {Hartstein}}, \bibinfo {author} {\bibfnamefont
  {M.}~\bibnamefont {Kiourlappou}}, \bibinfo {author} {\bibfnamefont
  {A.}~\bibnamefont {Srivastava}}, \bibinfo {author} {\bibfnamefont {M.~D.}\
  \bibnamefont {Johannes}}, \bibinfo {author} {\bibfnamefont {T.~P.}\
  \bibnamefont {Murphy}}, \bibinfo {author} {\bibfnamefont {J.-H.}\
  \bibnamefont {Park}}, \bibinfo {author} {\bibfnamefont {L.}~\bibnamefont
  {Balicas}}, \bibinfo {author} {\bibfnamefont {G.~G.}\ \bibnamefont
  {Lonzarich}}, \bibinfo {author} {\bibfnamefont {G.}~\bibnamefont
  {Balakrishnan}},\ and\ \bibinfo {author} {\bibfnamefont {S.~E.}\ \bibnamefont
  {Sebastian}},\ }\href {https://doi.org/10.1126/science.aaa7974} {\bibfield
  {journal} {\bibinfo  {journal} {Science}\ }\textbf {\bibinfo {volume}
  {349}},\ \bibinfo {pages} {287} (\bibinfo {year} {2015})}\BibitemShut
  {NoStop}%
\bibitem [{\citenamefont {Han}\ \emph {et~al.}(2019)\citenamefont {Han},
  \citenamefont {Li}, \citenamefont {Zhang}, \citenamefont {Sullivan},\ and\
  \citenamefont {Du}}]{hanAnomalousConductanceOscillations2019}%
  \BibitemOpen
  \bibfield  {author} {\bibinfo {author} {\bibfnamefont {Z.}~\bibnamefont
  {Han}}, \bibinfo {author} {\bibfnamefont {T.}~\bibnamefont {Li}}, \bibinfo
  {author} {\bibfnamefont {L.}~\bibnamefont {Zhang}}, \bibinfo {author}
  {\bibfnamefont {G.}~\bibnamefont {Sullivan}},\ and\ \bibinfo {author}
  {\bibfnamefont {R.-R.}\ \bibnamefont {Du}},\ }\href
  {https://doi.org/10.1103/PhysRevLett.123.126803} {\bibfield  {journal}
  {\bibinfo  {journal} {Physical Review Letters}\ }\textbf {\bibinfo {volume}
  {123}},\ \bibinfo {pages} {126803} (\bibinfo {year} {2019})}\BibitemShut
  {NoStop}%
\bibitem [{\citenamefont {Xiao}\ \emph {et~al.}(2019)\citenamefont {Xiao},
  \citenamefont {Liu}, \citenamefont {Samarth},\ and\ \citenamefont
  {Hu}}]{xiaoAnomalousQuantumOscillations2019}%
  \BibitemOpen
  \bibfield  {author} {\bibinfo {author} {\bibfnamefont {D.}~\bibnamefont
  {Xiao}}, \bibinfo {author} {\bibfnamefont {C.-X.}\ \bibnamefont {Liu}},
  \bibinfo {author} {\bibfnamefont {N.}~\bibnamefont {Samarth}},\ and\ \bibinfo
  {author} {\bibfnamefont {L.-H.}\ \bibnamefont {Hu}},\ }\href
  {https://doi.org/10.1103/PhysRevLett.122.186802} {\bibfield  {journal}
  {\bibinfo  {journal} {Physical Review Letters}\ }\textbf {\bibinfo {volume}
  {122}},\ \bibinfo {pages} {186802} (\bibinfo {year} {2019})}\BibitemShut
  {NoStop}%
\bibitem [{\citenamefont {Shao}\ and\ \citenamefont
  {Dai}(2023)}]{shaoElectricalBreakdownExcitonic2023}%
  \BibitemOpen
  \bibfield  {author} {\bibinfo {author} {\bibfnamefont {Y.}~\bibnamefont
  {Shao}}\ and\ \bibinfo {author} {\bibfnamefont {X.}~\bibnamefont {Dai}},\
  }\href@noop {} {\bibinfo {title} {Electrical {{Breakdown}} of {{Excitonic
  Insulator}}}},\ \bibinfo {howpublished} {https://arxiv.org/abs/2302.07543v1}
  (\bibinfo {year} {2023})\BibitemShut {NoStop}%
\bibitem [{\citenamefont {Brown}(1964)}]{Brown1964}%
  \BibitemOpen
  \bibfield  {author} {\bibinfo {author} {\bibfnamefont {E.}~\bibnamefont
  {Brown}},\ }\bibfield  {journal} {\bibinfo  {journal} {Physical Review}\
  }\textbf {\bibinfo {volume} {133}},\ \href
  {https://doi.org/10.1103/PhysRev.133.A1038} {10.1103/PhysRev.133.A1038}
  (\bibinfo {year} {1964})\BibitemShut {NoStop}%
\end{thebibliography}%
\bibliographystyle{apsrev4-2}


\end{document}